\documentclass[aps,10pt,nofootinbib]{revtex4-2}
\usepackage{amsmath}
\usepackage{amssymb}
\usepackage{booktabs}
\usepackage{graphicx}
\usepackage{hyperref}
\usepackage{float}
\usepackage[x11names,dvipsnames]{xcolor}
\usepackage{tikz}
\usetikzlibrary{arrows.meta,positioning,calc,fit,backgrounds,decorations.pathreplacing}

\begin{document}

\title{Emergent Space, Time, and Lorentz Symmetry from Binary Sequences}

\author{Sam Powers and Dejan Stojkovic}
\affiliation{HEPCOS, Department of Physics, SUNY at Buffalo, Buffalo, NY 14260, USA}

\begin{abstract}
We construct an information theory framework in which the fundamental objects are binary
sequences of length $n$, equipped with the bitwise XOR operation. The only physical
observables are counts of XOR-generated symbol classes, while the exact
locations of symbols are inaccessible. Averaging over those
locations ultimately yields the Minkowski interval as an invariant object that maximizes (information) entropy. 
Correlations between two binary sequences are base-4 sequences that we label as ``events'', and events are connected with maps (which are also base-4 sequences). The entire kinematic structure of Special Relativity is recovered under minimal assumptions, i.e. counts that belong to the maps and represent space and time increments carry equal informational weight. The central claim of the paper is that Lorentz
symmetry is the typical large-$n$ behaviour of XOR counting. At finite $n$
the framework yields a discrete rapidity spectrum, a single-map bound
$\gamma_{\max}=O(\sqrt{n})$, and interval fluctuations of relative size
$O(n^{-1/2})$, with standard special relativity recovered as $n\to\infty$. The velocity composition law holds typically, with
exponentially rare exceptions. However, the light cone and one null coordinate are exact
for every microscopic configuration, so the symmetry group is undeformed and dispersion
relations are unmodified. Thus, the theory, though discrete, 
implies no Lorentz violation of the standard phenomenological kind. Ultra-high-energy cosmic rays already
require $n\gtrsim10^{23}$; interferometry excludes the variant in which $n$
scales linearly with system size, leaving an area law that predicts a
universal length uncertainty of exactly one Planck length. The most striking prediction of the framework concerns systems at the maximum
of their information capacity, where $n$ is necessarily finite and the
deviations from the standard spacetime description become large: black
holes and de Sitter space. There the corrections are of order one within a
Planck proper length of the horizon, regardless of the horizon's size, and
the spacetime description fails altogether at the endpoint of black-hole
evaporation, where $n$ itself is of order unity.
\end{abstract}
\maketitle

\section{Introduction}

The standard formulation of special relativity begins with a pre-existing
spacetime manifold endowed with the Lorentz group as a spacetime symmetry.
Coordinates, intervals, and inertial frames are introduced from the outset.
The philosophy adopted here is different. We begin with no spacetime at all.
The primitive objects are binary sequences of length $n$. The primitive operation is XOR.
The primitive observables are counts. The question is then simple:

\begin{quote}
\emph{What geometry is seen by observers whose access to the underlying binary
world is restricted to count information?}
\end{quote}

If we give equal informational weight to the counts, the answer is simple
\begin{quote}
\textbf{Lorentz symmetry is the typical large-$n$ behaviour of XOR counting.}
\end{quote}
We say \emph{typical}, because the
symmetry holds for almost all configurations rather than for all of them, and
\emph{large-$n$}, because the corrections are controlled by the sequence length
and vanish only in the limit. And \emph{counting}, because the exact
microscopic theory is not Lorentzian at all: it is an abelian theory of subsets 
under symmetric difference (XOR), and the Lorentz group appears only after one
projects subsets onto their cardinalities.

The question is motivated by earlier work on binary-sequence descriptions of
quantum systems, in which observable physics emerged not from individual
sequences but from equivalence classes characterised only by counts of symbols that appear with sequences. In the previous work \cite{Powers:2021rfg}, an alternative way to modeling spin was defined. Among the other things, quantum mechanical rules for angular momentum composition and Clebsch-Gordan coefficients were derived, and the standard quantum mechanical results were recovered in the limit of the infinite length sequences.  The main message was that particles, viewed as states with definite angular momentum quantum numbers, are actually correlations between binary sequences (and therefore they do not even exist in between two correlations/measurements). This  work was then extended in \cite{Powers:2023ydg}, where an analog of the Wigner's d-matrix formula was derived and applied to two sequential Stern-Gerlach experiments. Along the way, a completely new approach to quantum mechanics was developed, without the Schrodinger equation, Born rule, and a measurement problem \cite{PowersThesis2023,Powers:2023sqf,Berglund:2022kew,QHO}. In addition, the whole group structure of the Standard Model of particle physics can be represented by discrete sequences \cite{SM}, in the form of which many salient features of the Standard Model become visible and allow long-standing problems to be attacked from a completely new perspective.

The central observation of the present work is that the same counting principle
generates a causal structure, and that Lorentz geometry appears as the
universal effective geometry associated with it. The logic of the construction
is
\[
\text{XOR}\;\rightarrow\;\text{counts}\;\rightarrow\;\text{count flow}
\;\rightarrow\;\text{causal cone}\;\rightarrow\;\text{Minkowski metric}
\;\rightarrow\;\text{Lorentz group}.
\]
This ordering matters. The Lorentz group, the metric, and the light cone are
not assumed; spacetime geometry emerges only after coarse-graining under minimal assumptions.

The XOR group on length-$n$ sequences is the elementary abelian group
$\mathbb{Z}_2^{\,n}$, in which every element is its own inverse, whereas the
boosts form the non-compact group $SO(1,1)\cong(\mathbb{R},+)$, in which no
non-identity element is an involution. No isomorphism between them exists. 
The coarse-grained formulation avoids that difficulty by never requiring it.
The XOR algebra generates a \emph{flow on count space}; Lorentz symmetry is a
property of that flow, not of the algebra --- the same distinction as between
a microscopic reversible dynamics and the hydrodynamic equations describing
its coarse-grained behaviour. This shift provides four things. First, the light
cone becomes a theorem rather than a postulate: it is the eigenvector
structure of the mixing matrix, and it is $\epsilon$-independent, so all
observers agree on it. Second, the conformal factor that must be removed to
obtain a pure Lorentz transformation acquires a physical meaning as the entropy
production of the coarse-graining, rather than being discarded by hand.
Third, the finite-$n$ corrections become calculable predictions. Fourth, the rapidity composition law acquires
the right logical status: relativistic velocity addition is a law of
large numbers rather than an algebraic identity, which is precisely what one
should expect of an emergent symmetry.

Because the framework claims to derive structures usually postulated, we are
explicit about the assumptions.
Section~\ref{sec:assumptions} lists the inputs, the outputs, the conventions,
and the open problems separately, and Section~\ref{sec:discussion} discusses
the most serious of the latter.

\subsection*{Relation to earlier work}

Derivations of the Lorentz transformation that do not postulate a speed of light 
principle have a long history, beginning with
Ignatowsky~\cite{Ignatowsky1910} and with Frank and
Rothe~\cite{FrankRothe1911}, and continuing through
L\'evy-Leblond~\cite{LevyLeblond1976} and many others~\cite{Pal2003}. These
derivations assume homogeneity, isotropy, and the group property of inertial
transformations, and obtain the Lorentz group with an undetermined invariant
speed. The present work belongs to that tradition in spirit but differs in
its starting point: those derivations begin with frames and velocities,
whereas here there are no frames, no velocities, and no coordinates at the
outset. What we add is not a weaker set of assumptions about spacetime, but
the absence of a spacetime to make assumptions about.

The framework also sits within the tradition of taking discrete or
informational structures as prior to spacetime. Wheeler's ``it from bit''
proposed that physical existence derives from binary registrations, and the
present construction is a literal attempt to carry that out for relativistic
kinematics. Causal set theory~\cite{Bombelli1987,Surya2019} shares our
conclusion that causal structure is prior to metric structure, and the
observation that a causal order plus a volume element recovers a Lorentzian
metric is similar to our logic in Section~\ref{sec:metric}. There are however
two important differences. Causal sets postulate the causal order and derive
the metric, whereas here the causal cone is itself derived, as the
$\epsilon$-independent eigenstructure of a counting flow. Also, in causal set
theory Lorentz invariance is not broken by the discreteness but is maintained
statistically, through random (Poisson) sprinkling, at the price of radical
nonlocality~\cite{Surya2019}; the present framework preserves it by
construction and must instead work to produce observable consequences.

\section{Binary Sequences}
\label{sec:sequences}

The fundamental objects are pairs (or correlations) of binary sequences defining an event $e$
\[
e=(r,s),\qquad r,s\in\{0,1\}^n .
\]
The two registers ($r$ for a referent sequence and $s$ for the system reference) are interpreted operationally as a clock register and a ruler
register. At each position one finds one of four symbols
\[
A=(0,0),\qquad C=(1,0),\qquad D=(0,1),\qquad B=(1,1).
\]
Thus, correlations between two base-2 sequences and base-4 sequences, with the basis $A,B,C,D$.
The collection $\{A,B,C,D\}$ forms the Klein four-group under XOR operation $\oplus$ (i.e. addition modulo two), which acts
independently on the two registers,
\[
(r_a,s_a)\oplus(r_b,s_b)=(r_a\oplus r_b,\;s_a\oplus s_b),
\]
where subscripts $a$ and $b$ refer to the events $a$ and $b$.
The full composition table is
\[
\begin{array}{c|cccc}
\oplus & A & C & D & B \\\hline
A      & A & C & D & B \\
C      & C & A & B & D \\
D      & D & B & A & C \\
B      & B & D & C & A
\end{array}
\]
This microscopic algebra contains no
spacetime, no metric, and no Lorentz symmetry.

\section{The Counting Postulate}
\label{sec:counting}

The key physical assumption of the framework is that the ordered location of
symbols is unobservable. Only the total counts (hats over the symbols)
\[
(\hat A,\hat B,\hat C,\hat D),\qquad \hat A+\hat B+\hat C+\hat D=n,
\]
carry physical significance: all sequences possessing the same count vector
represent the same physical state. This is the same postulate under which
quantum indeterminacy was obtained in earlier work in this programme, and we
import it unchanged. It is a genuine physical assumption, not a mathematical
convenience: it asserts that the microscopic ordering is not merely unknown to
a particular observer but is not part of the state.

For fixed counts the number of microscopic realisations is the multinomial
coefficient counting permutations with repetitions
\begin{equation}
\Omega=\frac{n!}{\hat A!\,\hat B!\,\hat C!\,\hat D!},
\label{eq:omega}
\end{equation}
and for large $n$, by Stirling's formula,
\begin{equation}
\ln\Omega = nH(P)+O(\ln n),\qquad
P_i=\frac{\hat i}{n},\qquad
H(P)=-\sum_i P_i\ln P_i .
\label{eq:entropy}
\end{equation}
where a hat over the symbol, ${\hat i}$, denotes how many times a particular symbol appears in a sequence.
The natural configuration space of the theory is therefore not the space of
$4^n$ sequences but the space of count distributions, which for large $n$ is
the three-dimensional probability simplex. Each point of that simplex carries
a statistical weight $\Omega$, and hence an entropy. Note that is information entropy (Shannon), not physical entropy. This is the first step
toward geometry, and we return to it in Section~\ref{sec:entropy}.

\section{Observer-Independent Maps}
\label{sec:maps}

Two events $e_a$ and $e_b$, are related to each other via the \emph{worldline map}
\begin{equation}
M_{ab}=e_a\oplus e_b ,
\label{eq:map}
\end{equation}
which is also a base-4 sequence with its own counts.
A change of observer corresponds to XORing every event by that observer's
reference sequence: Alice assigns to the event $e$ the coordinates
$\tilde e^{\,A}=e^A\oplus e$, and Bob assigns $\tilde e^{\,B}=e^B\oplus e$.
Because XOR is involutive,
\begin{equation}
(x\oplus y)\oplus(x\oplus z)=y\oplus z ,
\label{eq:cancel}
\end{equation}
all reference information cancels, and $M_{ab}$ is the same object for every
observer. This is the analogue of the statement that the separation between
two events is more primitive than the coordinates of either. 
Note also that Alice and Bob agree about the worldline as a set of events connected by maps, $M_{ab}$, between them, but they do not have to agree about the coordinate assignments (values) they give to these events.  

Two consequences are worth recording immediately; both are exact and neither
requires coarse-graining. First, maps are invariant under a common translation
of both endpoints:
\begin{equation}
(e_a\oplus\tau)\oplus(e_b\oplus\tau)=e_a\oplus e_b ,
\label{eq:transinv}
\end{equation}
exactly as $\Delta x^\mu$ is translation-invariant in ordinary relativity
while $x^\mu$ is not. We will need this in Section~\ref{sec:3plus1} to place
the inhomogeneous part of the Poincar\'e transformation where it belongs.
Second, for any three observers the maps obey an exact cocycle,
\begin{equation}
M_{ac}=M_{ab}\oplus M_{bc},
\label{eq:cocycle}
\end{equation}
again by Eq.~\eqref{eq:cancel}. This  identity has sharp
consequences for the composition of frame changes
(Section~\ref{sec:velocity}).

At the macroscopic level every map is described by its count vector
$(\hat A_M,\hat B_M,\hat C_M,\hat D_M)$, and nothing else about it is physical.

\section{Emergent Displacement Space}
\label{sec:displacement}

Since the four symbols, $A,B,C,D$, are on an equal footing as group elements, the group structure alone does not distinguish them,
but the \emph{product structure} does. An event is by construction a pair of
operationally distinct registers, while the map acts on them, so relative to that decomposition:

\begin{itemize}
\item $A_M=(0,0)$ is the identity and encodes no change in either register, so it
cannot label a direction of displacement;
\item $B_M=(1,1)$ changes both registers simultaneously and is not independent,
being $B_M=C_M\oplus D_M$; it is a mixing element;
\item $C_M=(1,0)$ and $D_M=(0,1)$ are the unique symbols changing exactly one
register each.
\end{itemize}

Accordingly, the symbol counts of $M_{ab}$ will ultimately give spacetime displacements as
\begin{equation}
  \Delta t = \hat{C}_{M},\qquad \Delta x = \hat{D}_{M},
  \label{eq:spacetimedict}
\end{equation}
while the count $\hat{B}_{M}$ will give boosts.

The automorphisms of the Klein group that preserve the register decomposition
therefore reduce to the single exchange $C\leftrightarrow D$, which swaps the
labels ``clock'' and ``ruler'' and reverses the signature convention
$(+,-)\to(-,+)$. Which register is called the clock is fixed by the time
ordering requirement of Section~\ref{sec:assumptions}. The identification is
thus forced up to that one convention.

So up to this point we only know that the displacement information is carried by the counts
\[
\hat C_M,\qquad \hat D_M,
\]
assembled into
\begin{equation}
X=\begin{pmatrix} \hat C_M\\ \hat D_M\end{pmatrix},
\label{eq:X}
\end{equation}
a point of a two-dimensional macroscopic count space. No spacetime
interpretation is attached to $X$ at this stage: $\hat C_M$ counts positions where
the clock register changed and the ruler register did not, $\hat D_M$ counts the
reverse. Whether they behave like a time and a space displacement is a
question to be settled by the dynamics, not by naming them.

\section{Frame Changes as Count Reclassification}
\label{sec:frame}

We saw that the worldline map $M_{ab}$ connects two events on the worldine. We now define an inter-frame map $\beta$ which connects two different observers, i.e. their frames. The inter-frame map between Alice and Bob is
\begin{equation}
 \beta=(r^A\oplus r^B,\,s^A\oplus s^B) .
 \end{equation}
  Recall that we do not know the exact composition of the map, only the number of times a certain element appears in it.
Let two observers differ by an inter-frame map $\beta$ carrying $\hat B_\beta$
symbols of type $B$, and write
\begin{equation}
\epsilon=\frac{\hat B_\beta}{n}
\label{eq:epsilon}
\end{equation}
for the fraction of positions at which both bits are flipped. At such a
position the roles of clock and ruler are exchanged, $C\leftrightarrow D$;
elsewhere they are not. A frame change is therefore not a rotation of axes but
a \emph{reclassification of counts}.

Before averaging, let us write the re-reading exactly. Let $k$ be the number of
clock positions of $M$ that coincide with a $B$ of $\beta$, and $l$ the number
of ruler positions that do. Writing $M'$ for the re-read map,
\begin{equation}
\hat C_{M'}=\hat C_M-k+l,\qquad \hat D_{M'}=\hat D_M-l+k .
\label{eq:exact}
\end{equation}
This is exact for every microscopic configuration, and it is linear in the
counts, a property inherited from the additivity of counting over positions
rather than assumed.

In the combinations
\begin{equation}
u=\hat C_M+\hat D_M,\qquad w=\hat C_M-\hat D_M,
\label{eq:uw}
\end{equation}
Eq.~\eqref{eq:exact} splits into two statements of opposite character. Adding
gives $u'=u$
%\begin{equation}
%u'=u
%\label{eq:uexact}
%\end{equation}
\emph{identically}, for every placement of the boost symbols and every finite
$n$, with no averaging whatsoever: the exchange permutes clock and ruler
positions without creating or destroying either. Subtracting gives $w'=w-2(k-l)$
%\begin{equation}
%w'=w-2(k-l),
%\%label{eq:wexact}
%\end{equation}
which does depend on the microscopic configuration. One of the two null
coordinates is thus a strict combinatorial invariant, and the entire
configuration dependence of the frame change is concentrated in the other.

\subsection{The coarse-grained flow}

Under the counting postulate the positions of the $B$ symbols carry no physical
information, so all placements are weighted equally. Then $k$ and $l$ are
multivariate hypergeometric, with
\[
\langle k\rangle=\epsilon\,\hat C_M,\qquad \langle l\rangle=\epsilon\,\hat D_M,
\]
and the mean transformation of the count vector is
\begin{equation}
X'=S(\epsilon)X,\qquad
S(\epsilon)=\begin{pmatrix}1-\epsilon & \epsilon\\ \epsilon & 1-\epsilon\end{pmatrix}.
\label{eq:S}
\end{equation}
This matrix is forced by count averaging: each clock
count is re-read as a ruler count with probability $\epsilon$ and each ruler
count as a clock count with the same probability, and there is nothing else it
could be.

Three properties of $S(\epsilon)$ will do all the work below. It is symmetric;
it has equal diagonal entries; and it is doubly stochastic. The first two are
consequences of the equal-weight treatment of the two registers, the third of
the fact that re-reading neither creates nor destroys positions.

\subsection{Fluctuations about the mean}
\label{sec:fluct}

Because the flow \eqref{eq:S} is a mean, the theory makes a definite prediction
about the spread around it, and this is the principal quantitative signature of
finite $n$. From the hypergeometric moments
\[
\mathrm{Var}(k)=\mathcal F\,\frac{\hat C_M}{n}\Bigl(1-\frac{\hat C_M}{n}\Bigr),\quad
\mathrm{Var}(l)=\mathcal F\,\frac{\hat D_M}{n}\Bigl(1-\frac{\hat D_M}{n}\Bigr),\quad
\mathrm{Cov}(k,l)=-\mathcal F\,\frac{\hat C_M}{n}\frac{\hat D_M}{n},
\]
with the finite-population factor
\[
\mathcal F=\frac{\hat B_\beta\,(n-\hat B_\beta)}{n-1},
\]
and using $w'=w-2(k-l)$, the cross terms combine into a perfect square and
\begin{equation}
\mathrm{Var}(w')=4\,\mathcal F\left(\frac{u}{n}-\frac{w^2}{n^2}\right).
\label{eq:varw}
\end{equation}
The covariance is negative because a position spent on a clock site is one
unavailable to a ruler site. Three checks confirm the formula: it vanishes at
$\hat B_\beta=0$ and $\hat B_\beta=n$, where the re-reading is deterministic;
it is non-negative, since $w^2\le u^2\le un$; and it is largest at
$\hat B_\beta=n/2$, where the placement is least constrained. We have also
verified it against direct enumeration and Monte Carlo sampling.

Equality in $w^2\le u^2\le un$ requires $|w|=u=n$, that is $\hat D_M=0$ and $\hat C_M=n$.
For such a maximal ``rest'' map every $B$ symbol necessarily lands on a clock
position, so $k=\hat B_\beta$ and $l=0$ with no freedom at all, and the
re-reading is exactly deterministic. This will matter in
Section~\ref{sec:continuum} when we ask what it would take to observe the
discreteness.

Since $ds^2=uw$ and $u$ is exactly invariant, the fluctuation of the interval
is carried entirely by $w$:
\begin{equation}
\mathrm{Var}(ds^2)=u^2\,\mathrm{Var}(w').
\label{eq:vards}
\end{equation}
At fixed symbol fractions $\mathrm{Var}(w')=O(n)$ against $\langle w'\rangle=O(n)$,
so the fractional spread falls as $n^{-1/2}$ and the coarse-grained description
becomes sharp in the continuum limit.

\section{The Emergent Causal Cone}
\label{sec:cone}

We can now state the central structural result. The matrix $S(\epsilon)$ has
eigenvectors
\[
u=\hat C_M+\hat D_M,\qquad w=\hat C_M-\hat D_M,
\]
with eigenvalues
\[
1\qquad\text{and}\qquad 1-2\epsilon .
\]
The remarkable fact is that \emph{the eigenvectors do not depend on $\epsilon$}.
Every observer, whatever the value of $\hat B_\beta$ relating them to any
other, identifies the same two distinguished directions in count space. This
follows from the structural properties noted above: a symmetric matrix with
equal diagonal entries has eigenvectors $(1,1)$ and $(1,-1)$ regardless of its
off-diagonal entry.

The Lorentzian signature is therefore \emph{not} a consequence of the XOR
algebra alone, and we should not claim more than the argument gives. The Klein
group by itself does not distinguish the pair $(C,D)$ from any other pair of
non-identity elements. What produces the null directions at $45^\circ$, and
hence the signature $(+,-)$, is the conjunction of two inputs from
Section~\ref{sec:inputs}: the product structure (i), which singles out
$C$ and $D$ as the single-register symbols, and equal information weight
(iii), which makes the diagonal entries of $S$ equal. Relax (iii) --- weight
the registers differently --- and the eigenvectors rotate away from
$45^\circ$ and the invariant speed takes a different value. What the counting
supplies on its own is that there \emph{is} an $\epsilon$-independent
eigenbasis, and hence an observer-independent cone; that this cone has the
Minkowski signature follows from (i) and (iii).

The set
\begin{equation}
w=0 \quad\Longleftrightarrow\quad \hat C_M=\hat D_M
\label{eq:cone}
\end{equation}
is therefore mapped to itself by every frame change in the family, exactly and
at every finite $n$. The binary counting theory thus produces an invariant
causal structure before any metric has been introduced. Nothing in the
construction of $S$ refers to light, to causality, or to a maximum speed; the
cone is the fixed locus of the mixing flow and nothing more. This is the
counting analogue of the Ignatowsky-type observation~\cite{Ignatowsky1910}
that an invariant speed
follows from the structure of the transformations rather than from a light
postulate, with the role of the group property played by the eigenvector
structure of the coarse-grained flow.

One caveat: the cone is invariant as a set, and so is each of its two sides,
$w>0$ and $w<0$; the flow does not by itself single out which side is
physical. Selecting the interior is nevertheless not an additional input:
Section~\ref{sec:causal} derives both the absence of superluminal frames and
the timelike, time-ordered character of worldlines from the operational
clock--ruler distinction of input (i) of Section~\ref{sec:inputs}, together
with the definition of a worldline as the history of a system that ticks.
What survives as an input is persistence alone, input (iv) of
Section~\ref{sec:inputs}. On the other hand, the counts are non-negative
by construction, so $u=\hat C_M+\hat D_M\ge0$ automatically; since $u$ is exactly invariant,
the theory has no past-directed sector at all, and time orientability comes
for free rather than as an extra assumption.

\section{Emergence of the Minkowski Metric}
\label{sec:metric}

The invariant cone determines a quadratic form essentially uniquely. Suppose
$Q(\hat C_M,\hat D_M)$ is a quadratic form vanishing on the cone. Writing $Q$ in the
eigenbasis $(u,w)$ as $Q=c_1 u^2+c_2 uw+c_3 w^2$ and imposing $Q=0$ on
$w=0$ for all $u$ gives $c_1=0$; imposing it on the second branch
of the cone removes the $w^2$ term. Hence
\begin{equation}
Q=c_2\,uw ,
\label{eq:Q}
\end{equation}
and $c_2$ is an overall normalisation. Substituting the definitions,
\begin{equation}
Q\propto (\hat C_M+\hat D_M)(\hat C_M-\hat D_M)=\hat C_M^2-\hat D_M^2 ,
\label{eq:interval}
\end{equation}
and we define
\begin{equation}
ds^2=\hat C_M^2-\hat D_M^2 ,
\label{eq:ds2}
\end{equation}
the Minkowski interval with signature $\eta=\mathrm{diag}(+1,-1)$.

The logic is the reverse of the usual one. One normally postulates a metric
and derives the light cone as its null set. Here the cone comes first, as the
eigenvector structure of a mixing matrix, and the metric is reconstructed from
it. The signature is not chosen: it is forced by the requirement that the form
vanish on a cone with two distinct real branches, which is precisely what a
symmetric mixing matrix with equal diagonal entries produces. Had the flow
instead had complex eigenvectors, the invariant form would have been positive
definite and no causal structure would have emerged.

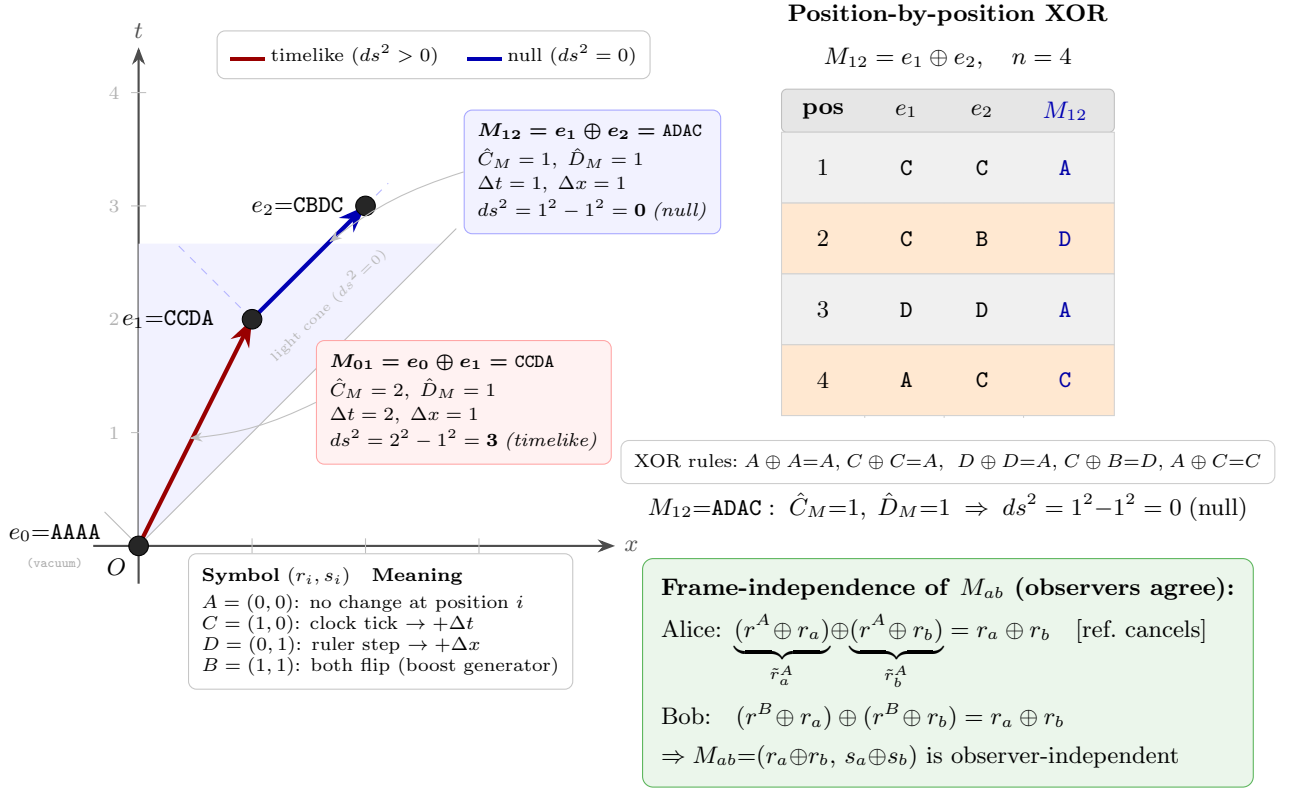
\begin{figure}[!htbp]
\centering
\begin{tikzpicture}[
  >=Stealth,
  event/.style={circle, draw=black, fill=black!85, inner sep=2.6pt},
  seq/.style={font=\small\ttfamily, align=center},
  axlab/.style={font=\small\itshape},
  cone/.style={gray!55, thin},
  nulledge/.style={->, thick, blue!70!black, line width=1.6pt},
  tledge/.style={->, thick, red!60!black, line width=1.6pt},
  mapbox/.style={draw=gray!60, fill=white, rounded corners=3pt,
                 inner sep=5pt, font=\scriptsize, align=left},
  thickax/.style={->, thick, black!70},
  rowA/.style={fill=gray!12},
  rowB/.style={fill=orange!18},
]

%──────────────────────────────────────────────────────────────────────────────
% LEFT PANEL: Spacetime diagram
% Events: e0=(x=0,t=0), e1=(x=1,t=2), e2=(x=2,t=3) in (x,t) units; 1 unit = 1.5cm
%──────────────────────────────────────────────────────────────────────────────
\def\sc{1.5}

% axes
\draw[thickax] (-0.6,0) -- (6.3,0) node[axlab,right] {$x$};
\draw[thickax] (0,-0.5) -- (0,6.6) node[axlab,above] {$t$};
\node[axlab,below left=2pt] at (0,0) {$O$};

% tick marks
\foreach \ty/\lb in {1/$1$,2/$2$,3/$3$,4/$4$}{
  \draw[thin,gray!50] (-0.10,\ty*\sc)--( 0.10,\ty*\sc);
  \node[font=\scriptsize,gray!60,left] at (-0.14,\ty*\sc) {\lb};
}
\foreach \tx/\lb in {1/$1$,2/$2$,3/$3$}{
  \draw[thin,gray!50] (\tx*\sc,-0.10)--(\tx*\sc, 0.10);
  \node[font=\scriptsize,gray!60,below] at (\tx*\sc,-0.14) {\lb};
}

% light-cone shading from origin
\fill[blue!5] (0,0) -- (4.0,4.0) -- (0,4.0) -- cycle;
\draw[cone] (0,0) -- (4.2,4.2);
\draw[cone] (0,0) -- (-0.45,0.45);
\node[rotate=44,font=\tiny,gray!55] at (2.5,3.2) {light cone ($ds^2\!=\!0$)};

% dashed light cone from e1
\draw[blue!30,dashed,thin] (\sc,2*\sc) -- (\sc+1.8,2*\sc+1.8);
\draw[blue!30,dashed,thin] (\sc,2*\sc) -- (\sc-1.0,2*\sc+1.0);

% worldline arrows
\draw[tledge] (0,0) -- (\sc,2*\sc);
\draw[nulledge] (\sc,2*\sc) -- (2*\sc,3*\sc);

% events
\node[event] (E0) at (0,0) {};
\node[event] (E1) at (\sc,2*\sc) {};
\node[event] (E2) at (2*\sc,3*\sc) {};

% event labels with base-4 sequences
\node[seq,left=11pt] at (E0)
  {$e_0{=}\texttt{AAAA}$\\\textcolor{gray!65}{\tiny(vacuum)}};
\node[seq,left=11pt] at (E1)
  {$e_1{=}\texttt{CCDA}$};
\node[seq,left=5pt] at (E2)
  {$e_2{=}\texttt{CBDC}$};

% map annotation: M01 (timelike, red)
\node[mapbox,fill=red!5,draw=red!35] (BOX01) at (4.3,1.9)
  {$\boldsymbol{M_{01}=e_0\oplus e_1=\texttt{CCDA}}$\\[3pt]
   $\hat{C}_{M}=2,\;\hat{D}_{M}=1$\\[1pt]
   $\Delta t=2,\;\Delta x=1$\\[2pt]
   $ds^2=2^2-1^2=\mathbf{3}$\;\textit{(timelike)}};
\draw[->,gray!50,thin] (BOX01.west)
  to[bend left=10] (0.45*\sc,0.95*\sc);

% map annotation: M12 (null, blue)
\node[mapbox,fill=blue!5,draw=blue!35] (BOX12) at (6.0,4.95)
  {$\boldsymbol{M_{12}=e_1\oplus e_2=\texttt{ADAC}}$\\[3pt]
   $\hat{C}_{M}=1,\;\hat{D}_{M}=1$\\[1pt]
   $\Delta t=1,\;\Delta x=1$\\[2pt]
   $ds^2=1^2-1^2=\mathbf{0}$\;\textit{(null)}};
\draw[->,gray!50,thin] (BOX12.west)
  to[bend right=10] (1.68*\sc,2.68*\sc);

% arrow-type legend
\node[draw=gray!45,fill=white,rounded corners=3pt,
      font=\scriptsize,align=left,inner sep=5pt] at (3.9,6.5)
  {\textcolor{red!60!black}{\rule{13pt}{1.5pt}}\;timelike ($ds^2>0$)
   \quad
   \textcolor{blue!70!black}{\rule{13pt}{1.5pt}}\;null ($ds^2=0$)};

% symbol glossary
\node[draw=gray!45,fill=white,rounded corners=3pt,
      font=\scriptsize,align=left,inner sep=5pt] at (3.2,-1)
  {\textbf{Symbol} $(r_i,s_i)$\quad\textbf{Meaning}\\[2pt]
   $A=(0,0)$: no change at position $i$\\
   $C=(1,0)$: clock tick $\to+\Delta t$\\
   $D=(0,1)$: ruler step $\to+\Delta x$\\
   $B=(1,1)$: both flip (boost generator)};

%──────────────────────────────────────────────────────────────────────────────
% RIGHT PANEL: Position-by-position XOR table + frame-independence
%──────────────────────────────────────────────────────────────────────────────
\begin{scope}[xshift=8.5cm]

% title
\node[font=\small\bfseries] at (2.2,4.68*\sc)
  {Position-by-position XOR};
\node[font=\small] at (2.2,4.30*\sc)
  {$M_{12}=e_1\oplus e_2,\quad n=4$};

% table header row
\draw[fill=gray!20,draw=gray!50,rounded corners=2pt]
  (0,3.65*\sc) rectangle (4.4,4.03*\sc);
\node[font=\small\bfseries] at (0.55,3.84*\sc) {pos};
\node[font=\small\bfseries] at (1.65,3.84*\sc) {$e_1$};
\node[font=\small\bfseries] at (2.65,3.84*\sc) {$e_2$};
\node[font=\small\bfseries,blue!70!black] at (3.75,3.84*\sc) {$M_{12}$};

% vertical dividers
\foreach \xd in {1.10, 2.15, 3.20}{
  \draw[gray!38,thin] (\xd,3.65*\sc) -- (\xd,1.13*\sc);
}

% data rows: e1=CCDA, e2=CBDC, M12=ADAC
% pos, e1, e2, M12, style
\foreach \row/\p/\ea/\eb/\em/\sty in {
  1/1/C/C/A/rowA,
  2/2/C/B/D/rowB,
  3/3/D/D/A/rowA,
  4/4/A/C/C/rowB}{
  \pgfmathsetmacro{\ybot}{(3.65 - \row*0.625)*\sc}
  \pgfmathsetmacro{\ytop}{(3.65 - (\row-1)*0.625)*\sc}
  \pgfmathsetmacro{\ymid}{(\ybot+\ytop)/2}
  \draw[\sty,draw=gray!35] (0,\ybot) rectangle (4.4,\ytop);
  \node[font=\small] at (0.55,\ymid) {\p};
  \node[font=\small\ttfamily] at (1.65,\ymid) {\ea};
  \node[font=\small\ttfamily] at (2.65,\ymid) {\eb};
  \node[font=\small\ttfamily,blue!65!black] at (3.75,\ymid) {\em};
}

% XOR rules used
\node[draw=gray!40,fill=white,rounded corners=3pt,
      font=\scriptsize,align=center,inner sep=5pt] at (2.2,0.74*\sc)
  {XOR rules:\;$A\oplus A{=}A$,\;$C\oplus C{=}A$,\;
   $D\oplus D{=}A$,\;$C\oplus B{=}D$,\;$A\oplus C{=}C$};

% result box
\node[font=\small,align=center] at (2.2,0.35*\sc)
  {$M_{12}{=}\texttt{ADAC}:\;
   \hat{C}_M{=}1,\;\hat{D}_M{=}1
   \;\Rightarrow\;ds^2=1^2{-}1^2=0\;(\text{null})$};

% frame-independence box
\node[draw=Green4!65,fill=Green4!8,rounded corners=4pt,
      font=\small,align=left,inner sep=7pt] at (2.2,-1.12*\sc)
  {\textbf{Frame-independence of }$M_{ab}$\textbf{\ (observers agree):}\\[4pt]
   Alice:\enspace
   $\underbrace{(r^A\!\oplus r_a)}_{\tilde r^A_a}
   \!\oplus\!
   \underbrace{(r^A\!\oplus r_b)}_{\tilde r^A_b}
   =r_a\oplus r_b$\quad[ref.\ cancels]\\[4pt]
   Bob:\quad
   $(r^B\!\oplus r_a)\oplus(r^B\!\oplus r_b)=r_a\oplus r_b$\\[4pt]
   $\Rightarrow M_{ab}{=}(r_a{\oplus}r_b,\,s_a{\oplus}s_b)$
   is observer-independent};

\end{scope}

% panel separator
%\draw[gray!28,dashed,thin] (7.7,-1.8) -- (7.7,6.8);

\end{tikzpicture}

\caption{%
\textbf{Left:} A worldline as a sequence of base-4 events in a spacetime
  Minkowski diagram ($t$ vertical, $x$ horizontal), for $n=4$ positions per
  sequence.
  Each event $e_i=\texttt{(symbol}_1\texttt{,symbol}_2\texttt{,symbol}_3\texttt{,symbol}_4\texttt{)}$ is a
  length-4 string over $\{A,B,C,D\}$; each symbol at position $i$ records
  the pair $(r_i,s_i)$ where $r_i$ and $s_i$ take values $0$ or $1$ representing a clock bit and ruler bit respectively.
  The map $M_{ab}=e_a\oplus e_b$ between two events is their bitwise XOR,
  computed position by position.
  Counting $C$-symbols gives $\Delta t$, counting $D$-symbols gives $\Delta x$,
  and the Minkowski interval is $ds^2=\hat{C}_{M}^2-\hat{D}_{M}^2$.
  The red arrow ($M_{01}$: $\hat{C}_M{=}2,\hat{D}_M{=}1$, $ds^2{=}3$) is
  a timelike step; the blue arrow ($M_{12}$: $\hat{C}_M{=}1,\hat{D}_M{=}1$,
  $ds^2{=}0$) is a null (lightlike) step lying exactly on the light cone of~$e_1$.
  \textbf{Right:} Explicit position-by-position XOR computation of
  $M_{12}=e_1\oplus e_2$, illustrating how each symbol in the map is
  determined by the Klein-four-group rule applied to the corresponding
  positions of $e_1$ and $e_2$.
  The frame-independence box shows that every observer's reference
  sequence cancels exactly from $M_{ab}$, so the map is a geometric object
  all observers agree on (Eq.~\protect\eqref{eq:cancel}).%
}
\label{fig:worldline}
\end{figure}

Figure~\ref{fig:worldline} collects the construction to this point in a
single picture: events as base-4 strings, maps as position-by-position XOR,
the counting dictionary \eqref{eq:spacetimedict}, the interval
\eqref{eq:ds2} with its timelike and null classes, and the exact
cancellation \eqref{eq:cancel} that makes every map observer independent.
The timelike step shown, $M_{01}=\texttt{CCDA}$ with $\hat C_M=2$,
$\hat D_M=1$, $ds^2=3$, is the same map worked through microscopically in
Appendix~\ref{app:n4}.

 Nothing in the counting fixes the constant $c_2$, and
nothing can: counts are dimensionless integers, and no function of integers
returns a unit of length. Choosing $c_2=1$ is a choice of the unit in which
intervals are measured. What the framework determines is the conformal class
of the metric together with its signature; the scale is a convention.

\section{Emergent Lorentz Symmetry}
\label{sec:lorentz}

Under the coarse-grained flow the interval is not invariant. Applying
Eq.~\eqref{eq:S},
\[
u'=u,\qquad w'=(1-2\epsilon)w,
\]
and therefore
\begin{equation}
ds'^2=(1-2\epsilon)\,ds^2 .
\label{eq:conformal}
\end{equation}
The fundamental count flow is thus \emph{conformal} rather than isometric: it
preserves the cone, hence all causal relations, but shrinks all intervals by a
common factor. Every conformal transformation factorises uniquely into a dilation and an
interval-preserving part,
\begin{equation}
S(\epsilon)=\sqrt{1-2\epsilon}\;\Lambda(\phi),
\label{eq:factor}
\end{equation}
where a short computation in the eigenbasis gives
$\Lambda=\mathrm{diag}\bigl(e^{\phi},e^{-\phi}\bigr)$ with
$e^{-\phi}=\sqrt{1-2\epsilon}$, that is
\begin{equation}
\phi=\frac12\ln\frac{1}{1-2\epsilon}
   =\frac12\ln\frac{n}{n-2\hat B_\beta}
   =\operatorname{artanh}\frac{\hat B_\beta}{n-\hat B_\beta}.
\label{eq:rapidity}
\end{equation}
In the original $(\hat C_M,\hat D_M)$ basis,
\begin{equation}
\Lambda(\phi)=\begin{pmatrix}\cosh\phi & \sinh\phi\\ \sinh\phi & \cosh\phi\end{pmatrix},
\label{eq:boost}
\end{equation}
which satisfies
\[
\det\Lambda=1,\qquad \Lambda^{T}\eta\,\Lambda=\eta,\qquad
\eta=\begin{pmatrix}1&0\\0&-1\end{pmatrix},
\]
so that $\Lambda\in SO(1,1)$. The positive off-diagonal entries are forced by
Eq.~\eqref{eq:factor}: counts mix additively in Eq.~\eqref{eq:S}, so
$\sqrt{1-2\epsilon}\,\Lambda(\phi)$ must have positive entries. The more
familiar form with $-\sinh\phi$ is $\Lambda(-\phi)$, and the choice between
the two is the direction-of-motion convention of
Section~\ref{sec:orientation}. Lorentz symmetry appears as the
interval-preserving subgroup of the universal count flow.

\subsection{The conformal factor is not physical}
\label{sec:whyconformal}

The factorisation \eqref{eq:factor} is mathematically unique, but uniqueness
of a decomposition does not by itself justify discarding half of it. It should
be said plainly that the coarse-grained flow \emph{as a whole} is not Lorentz
invariant: it is conformal, and identifying its interval-preserving factor
with Lorentz symmetry requires a physical argument for setting the dilation
aside. Three observations supply it, and the last one is the most important.

\paragraph{The dilation is worldline-independent.}
The factor $1-2\epsilon$ depends only on $\hat B_\beta$ and $n$, that is on the
frame map alone, and is identical for every map $M$ it acts on. It rescales
all intervals in the frame by one common number, leaving every ratio of
intervals unchanged. Since any measurement returns ratios, a uniform rescaling
of this kind is a change of unit rather than a physical deformation. A
transformation that rescaled different worldlines differently would be
detectable; this one is not.

\paragraph{The dilation does not affect causal structure.}
Because $S$ preserves the cone exactly, the causal relations among events, and
hence the entire conformal geometry, are already determined before the
factorisation is performed. What the conformal structure leaves undetermined
is exactly one scale, and this is what the dilation carries.

\paragraph{The dilation is the entropy production of the coarse-graining.}
This is the physically meaningful statement, developed in
Section~\ref{sec:entropy}. Because $S(\epsilon)$ is doubly stochastic, the
coarse-grained flow is irreversible: it strictly increases the entropy of the
count distribution unless $\epsilon=0$ or the state already lies on the cone.
The dilation factor $\sqrt{1-2\epsilon}<1$ is precisely the irreversible part
of the flow, and $\Lambda(\phi)$ is what remains after the entropy production
has been removed: invertible, interval-preserving, and reversible. Lorentz
symmetry, in this reading, is the \emph{reversible core} of an irreversible
flow --- the sense in which the framework is a statistical-mechanical theory
of relativity rather than an algebraic one, and the reason the factorisation
is a physical decomposition rather than a normalisation convention.

\section{The Rapidity Spectrum at Finite $n$}
\label{sec:spectrum}

Because $\hat B_\beta$ is an integer, Eq.~\eqref{eq:rapidity} yields a
\emph{discrete} set of accessible rapidities. Writing $\hat B_\beta=k$,
\begin{equation}
\phi_k=\frac12\ln\frac{n}{n-2k},\qquad
v_k=\tanh\phi_k=\frac{k}{n-k},
\label{eq:spectrum}
\end{equation}
with $k$ ranging over the integers for which the transformation is
sub-luminal; the range is
\begin{equation}
k=0,1,\dots,k_{\max},\qquad k_{\max}=\Bigl\lfloor\frac{n-1}{2}\Bigr\rfloor .
\label{eq:kmax}
\end{equation}
At $k=n/2$, possible only for even $n$, one has $v=1$ exactly: the frame
separation sits on the light cone. Sub-luminality is not an assumption here:
Section~\ref{sec:causal} shows, through the exact identity
\eqref{eq:complement}, that a re-reading at $k>n/2$
positions equals the re-reading at the complementary $n-k$ positions
composed with a global exchange of the registers, so the upper half of the
ladder is a re-description of the lower half, not new physics. Two features of this spectrum are worth
noting.

\paragraph{Spacing.} The gap between adjacent rungs is
\begin{equation}
\delta\phi_k=\phi_{k+1}-\phi_k=\frac12\ln\frac{n-2k}{n-2k-2}
\;\simeq\;\frac{1}{n-2k}
\label{eq:spacing}
\end{equation}
for $n-2k\gg1$. The ladder is finest near $v=0$, where the spacing is $1/n$,
and coarsens without bound as the light cone is approached, the last few rungs
being separated by amounts of order unity. The near-luminal region is thus the
most strongly quantised, which is the opposite of what a naive
``Planck-scale-only'' intuition would suggest.

\paragraph{A maximum single-map boost.} The largest accessible Lorentz factor
follows from $k=k_{\max}$:
\begin{equation}
\gamma_{\max}=\frac{n-k_{\max}}{\sqrt{n\,(n-2k_{\max})}}
=\begin{cases}
\dfrac{n+1}{2\sqrt n}\simeq\dfrac{\sqrt n}{2}, & n \text{ odd},\\[2ex]
\dfrac{n+2}{2\sqrt{2n}}\simeq\dfrac{\sqrt n}{2\sqrt 2}, & n \text{ even}.
\end{cases}
\label{eq:gammamax}
\end{equation}
The two cases differ by a factor $\sqrt2$ because the closest approach to the
cone is $n-2k_{\max}=1$ for odd $n$ but $2$ for even $n$. In either case
$\gamma_{\max}=O(\sqrt n)$: a \emph{single} frame map cannot encode an
arbitrarily large boost for finite $n$, because the information needed to specify a
near-luminal frame change grows without bound while the map has only $n$
positions. Larger boosts are reached by composing several maps, as described
next.

\section{The Velocity Addition Law}
\label{sec:velocity}

Suppose Alice and Bob differ by $\beta_1$ and Bob and Carol by $\beta_2$. By
the cocycle \eqref{eq:cocycle} the map relating Alice and Carol is
\begin{equation}
\beta_{\rm tot}=\beta_1\oplus\beta_2 ,
\label{eq:betatot}
\end{equation}
exactly. A position carries a $B$ in $\beta_{\rm tot}$ when it carries one in
exactly one of $\beta_1,\beta_2$: two exchanges at the same position cancel.
Hence
\begin{equation}
\hat B_{\rm tot}=\hat B_1+\hat B_2-2\,|\beta_1\cap\beta_2| .
\label{eq:Btot}
\end{equation}

For \emph{independently placed} boost symbols the expected overlap is
$\hat B_1\hat B_2/n$, so
\begin{equation}
\epsilon_{\rm tot}=\epsilon_1+\epsilon_2-2\epsilon_1\epsilon_2 ,
\label{eq:epstot}
\end{equation}
and therefore
\begin{equation}
1-2\epsilon_{\rm tot}=(1-2\epsilon_1)(1-2\epsilon_2).
\label{eq:multiplicative}
\end{equation}
The combination $1-2\epsilon$ is multiplicative, which by
Eq.~\eqref{eq:rapidity} means that rapidity is additive:
\begin{equation}
\phi_{\rm tot}=\phi_1+\phi_2 ,
\label{eq:phiadd}
\end{equation}
and the corresponding matrices satisfy
$\Lambda(\phi_1)\Lambda(\phi_2)=\Lambda(\phi_1+\phi_2)$. Taking $\tanh$ of
both sides gives the relativistic velocity addition law
\begin{equation}
v_{\rm tot}=\frac{v_1+v_2}{1+v_1v_2}.
\label{eq:vadd}
\end{equation}
Note also that $\epsilon_1,\epsilon_2<\tfrac12$ implies
$\epsilon_{\rm tot}<\tfrac12$: composing sub-luminal frame changes never
produces a super-luminal one, so the cone is respected by composition as well
as by individual transformations. (That every physical frame change has
$\epsilon<\tfrac12$ in the first place is derived in
Section~\ref{sec:causal}.)

\subsection{Is the calibration circular?}
\label{sec:notcircular}

An objection presents itself immediately: the rapidity \eqref{eq:rapidity}
was \emph{defined} so as to be additive, so recovering the relativistic
velocity-addition law is a tautology. It is not, because the two ingredients
are fixed independently and neither refers to the other.

\begin{itemize}
\item The calibration $\phi=-\tfrac12\ln(1-2\epsilon)$ is fixed by the
factorisation \eqref{eq:factor}: in the null basis $S=\mathrm{diag}(1,1-2\epsilon)$,
and removing the determinant to leave a unit-determinant factor gives
$\Lambda=\mathrm{diag}(e^{\phi},e^{-\phi})$ with $e^{-2\phi}=1-2\epsilon$. This
uses only the requirement that a frame change preserve $ds^2$. No composition
law enters.
\item The composition rule $\epsilon_{\rm tot}=\epsilon_1+\epsilon_2-2\epsilon_1\epsilon_2$
is fixed by the overlap statistics of subsets under symmetric difference. No
rapidity enters.
\end{itemize}

That these agree is therefore a genuine consistency check, and it could have
failed. To see that it could, hold the factorisation fixed --- so that $\phi$ is
what it is --- and ask whether $\phi$ would still be additive had the
combinatorics delivered some other composition rule:

\begin{center}
\begin{tabular}{lcc}
\toprule
hypothetical rule for $\epsilon_{\rm tot}$ & $\phi_{\rm tot}$ & additive? \\
\midrule
$\epsilon_1+\epsilon_2-2\epsilon_1\epsilon_2$ \ (XOR overlap) & $0.36699$ & \textbf{yes} \\
$\epsilon_1+\epsilon_2$ & $0.45815$ & no \\
$\epsilon_1+\epsilon_2-\epsilon_1\epsilon_2$ & $0.41049$ & no \\
$(\epsilon_1+\epsilon_2)/2$ & $0.17834$ & no \\
$\sqrt{\epsilon_1^2+\epsilon_2^2}$ & $0.29639$ & no \\
\bottomrule
\end{tabular}
\end{center}

\noindent
(evaluated at $\epsilon_1=0.20$, $\epsilon_2=0.10$, where
$\phi_1+\phi_2=0.36699$). Only the rule that XOR actually supplies makes the
factorisation-fixed $\phi$ additive, because only for that rule is $1-2\epsilon$
multiplicative. The calibration could not have been tuned to produce
additivity, since it was already determined before the composition law was
computed.

A related point applies to the velocity. One might suspect that
$v=\tanh\phi$ is imported from relativity. It is not: from
$e^{-2\phi}=1-2\epsilon$ one computes
\[
\tanh\phi=\frac{1-e^{-2\phi}}{1+e^{-2\phi}}=\frac{\epsilon}{1-\epsilon}
=\frac{\hat B_\beta}{n-\hat B_\beta},
\]
so the velocity is the ratio of boost symbols to non-boost symbols, a pure
counting ratio. That this ratio is the hyperbolic tangent of the additive
parameter is a result, not an identification. The input was the requirement that a frame change preserve $ds^2$,
which is what selects the unit-determinant factor in the first place, however we showed that the invariance of $ds^2$ comes from the XOR flow. Given
that requirement, the calibration is forced and its additivity is a theorem.

\subsection{Frame maps are subsets, not counts}
\label{sec:subsets}

The derivation above used the \emph{expected} overlap, and it is important to
see that this step is not trivial. For an actual triple of observers the
overlap is not an expectation but a definite number fixed by their three
reference sequences, and Eq.~\eqref{eq:betatot} is an exact identity rather
than a statistical statement.

The diagnosis is that $\hat B_\beta$ is not a complete invariant of a frame
map. The genuine object attached to $\beta$ is the \emph{set} of positions
carrying a boost symbol, $B$,
\begin{equation}
S_\beta\subset\{1,\dots,n\},\qquad |S_\beta|=\hat B_\beta ,
\label{eq:subset}
\end{equation}
and $\hat B_\beta$ is only its cardinality. Two frame maps with identical
counts but differently placed boost symbols compose differently, so the count
cannot be the frame coordinate; it is a projection of one.

At the level of subsets everything is exact. XOR composition of frame maps is
symmetric difference,
\begin{equation}
S_{AC}=S_{AB}\,\triangle\,S_{BC},
\label{eq:symdiff}
\end{equation}
and the family $\{S_\beta\}$ under $\triangle$ is precisely the group
$\mathbb{Z}_2^{\,n}$, in which the cocycle \eqref{eq:cocycle} holds
identically for every triple, with no averaging and no error term. The exact
theory is therefore a \emph{subset cocycle}, and this is the correct
microscopic statement of the composition law. Rapidity enters only through
\begin{equation}
\varrho(\beta)=1-\frac{2|S_\beta|}{n},\qquad
\phi=-\tfrac12\ln\varrho ,
\label{eq:varrho}
\end{equation}
which discards the placement and retains the cardinality. It is this
projection, and not XOR, that fails to be a homomorphism. The question is then
not whether the exact theory is consistent --- it is, trivially --- but whether
the projection to counts is \emph{asymptotically} a homomorphism.

\subsection{The emergent Lorentz cocycle}
\label{sec:cocycletheorem}

The precise statement is the following. Let $S_{AB}$ and $S_{BC}$ be
independently chosen uniform random subsets of
$\{1,\dots,n\}$ of sizes $\hat B_1=b_1n$ and $\hat B_2=b_2n$, with
$b_1,b_2$ fixed and the composed fraction
$\bar\epsilon=b_1+b_2-2b_1b_2$ bounded away from $\tfrac12$. Then the composed
rapidity obeys
\begin{equation}
\phi_{AC}=\phi_{AB}+\phi_{BC}+\Delta_n,
\label{eq:cocyclethm}
\end{equation}
where the discrepancy $\Delta_n$ has mean $O(n^{-1})$ and standard deviation
\begin{equation}
\sigma_\Delta=\frac{2}{1-2\bar\epsilon}
\sqrt{\frac{b_1b_2(1-b_1)(1-b_2)}{n}}\;=\;O\!\left(n^{-1/2}\right).
\label{eq:sigmadelta}
\end{equation}
Moreover, for any $\delta>0$,
\begin{equation}
\Pr\bigl(|\Delta_n|>\delta\bigr)\;\le\;2e^{-cn},\qquad c=c(\delta,b_1,b_2)>0 .
\label{eq:largedev}
\end{equation}
Relativistic rapidity addition therefore holds for almost all triples of
observers, with probability approaching one exponentially fast as
$n\to\infty$. The derivation --- an exercise in hypergeometric moments, error
propagation through the calibration, and Hoeffding's inequality for sampling
without replacement --- is given in Appendix~\ref{app:proofs}.

Monte Carlo sampling confirms every element of this statement: at $n=400$,
$b_1=0.2$, $b_2=0.3$ we find $\langle I\rangle=24.00$ and
$\mathrm{Var}(I)=13.50$ against the predicted $24.00$ and $13.47$; the product
$\sigma_\Delta\sqrt n$ is constant at $1.5275$ across $n=10^2$ to
$2.56\times10^4$; the bias satisfies $n\langle\Delta_n\rangle\to2.33$; and the
empirical large-deviation probabilities fall below the Hoeffding bound at
every $n$ tested. Appendix~\ref{app:cocycle} exhibits the count-level failure
with explicit, deliberately atypical subsets.

\subsection{Independence is not an additional assumption: it is the counting
measure applied to pairs}
\label{sec:maxent}

The composition law of Section~\ref{sec:cocycletheorem} was stated for
\emph{independently sampled} frame maps, and one might object that this is
an assumption about an ensemble of observers imported from outside. It is not, because a flat measure over joint placements and
statistical independence of the placements are the same assumption in
different words. The reply is that it is not an \emph{additional}
assumption: it is the flat measure of input (ii) of
Section~\ref{sec:inputs}, already adopted in Section~\ref{sec:counting},
applied to the pair of frame maps rather than to one, with the overlap
unobserved. What this section establishes is uniformity, i.e. one measure, used
everywhere, together with the rate function below, which quantifies how
departures from the typical overlap are suppressed under that measure.

One misreading should be forestalled: the framework does not treat symbol
arrangements as meaningful when discussing overlaps and as unobservable when
averaging. The arrangement is unobservable in both places; the overlap is not
a datum to be measured but a quantity to be averaged over, under the same
measure as the placement of boost symbols in a single frame change.

\paragraph{The flat measure induces the hypergeometric.}
By the counting postulate a physical state is specified by counts, with all
microscopic realisations equally weighted. For a pair of frame maps the
measurable data are the two cardinalities $\hat B_1=|S_{AB}|$ and
$\hat B_2=|S_{BC}|$; the microstates are all pairs of subsets with those
cardinalities. Counting them by overlap,
\begin{equation}
\#\bigl\{(S_1,S_2)\;:\;|S_1\cap S_2|=I\bigr\}
=\binom{n}{\hat B_1}\binom{\hat B_1}{I}\binom{n-\hat B_1}{\hat B_2-I},
\label{eq:jointcount}
\end{equation}
so that the flat measure on microstates induces
\begin{equation}
P(I)\;\propto\;\binom{\hat B_1}{I}\binom{n-\hat B_1}{\hat B_2-I},
\label{eq:hyperinduced}
\end{equation}
which is exactly the hypergeometric distribution assumed in
Section~\ref{sec:cocycletheorem}. The independence hypothesis is therefore
not a \emph{further} input: the flat measure on pairs with fixed
cardinalities is, word for word, the statement that the two placements are
independent given their sizes. The computation shows that using it here adds
nothing beyond input (ii) applied uniformly, and that refusing it would
amount to applying the postulate in one place and denying it in another.

\paragraph{Additivity is the entropy maximum.}
The joint count
\eqref{eq:jointcount} is maximised at
\begin{equation}
I^\star=\frac{\hat B_1\hat B_2}{n},
\label{eq:Istar}
\end{equation}
which is precisely the overlap at which rapidities add. In other words:
\begin{quote}
\emph{Relativistic velocity addition is the maximum-entropy composition law.}
\end{quote}
The overlap that makes rapidities additive is not a fine-tuned coincidence to
be imposed by hand; it is the overlap realised by the overwhelming majority of
microstates compatible with what an observer can measure. Deviations from
additivity are deviations from equilibrium.

\paragraph{The suppression rate, explicitly.}
This upgrades the exponential bound \eqref{eq:largedev} to an equality with a
computable rate. Writing $i=I/n$, $b_a=\hat B_a/n$, and using Stirling in
\eqref{eq:jointcount}, the entropy density of configurations with overlap
fraction $i$ is
\begin{equation}
\mathsf{s}(i)=b_1\,h\!\left(\frac{i}{b_1}\right)
+(1-b_1)\,h\!\left(\frac{b_2-i}{1-b_1}\right),
\qquad h(x)=-x\ln x-(1-x)\ln(1-x),
\label{eq:sdensity}
\end{equation}
and the probability of an atypical overlap obeys the large-deviation form
\begin{equation}
P(i)\;\asymp\;e^{-n\,J(i)},\qquad
J(i)=\mathsf{s}(i^\star)-\mathsf{s}(i),\qquad i^\star=b_1b_2 .
\label{eq:ratefn}
\end{equation}
$J$ is the entropy \emph{deficit} of the atypical configuration, so the
suppression of a violation of velocity addition is exactly the entropy cost of
producing it.

Two checks confirm that this is the same structure as the cocycle result of
Section~\ref{sec:cocycletheorem} rather than a parallel one. First, expanding
\eqref{eq:sdensity} about $i^\star$ gives
\begin{equation}
\mathsf{s}''(i^\star)=-\frac{1}{b_1b_2(1-b_1)(1-b_2)},
\qquad
J(i)\simeq\frac{(i-i^\star)^2}{2\,b_1b_2(1-b_1)(1-b_2)} .
\label{eq:Jquad}
\end{equation}
Second, propagating this through $|d\phi/di|=2/(1-2\bar\epsilon)$ reproduces
the standard deviation \eqref{eq:sigmadelta} exactly: at $b_1=0.2$, $b_2=0.3$
both give $\sigma_\Delta=0.076376$ at $n=400$ and $0.019094$ at $n=6400$. The
Gaussian core of the rate function is the fluctuation theorem; its tails are
the large-deviation bound. To make the numbers concrete, at the same
$(b_1,b_2)$ a discrepancy $\Delta\phi=-0.144$ carries $J=5.7\times10^{-3}$,
giving suppression $e^{-nJ}\approx0.10$ at $n=400$ but $\approx10^{-25}$ at
$n=10^4$; a discrepancy $\Delta\phi=-0.255$ carries $J=2.2\times10^{-2}$ and
is suppressed by $\sim10^{-97}$ at $n=10^4$.

\paragraph{A limitation of this argument.}
The maximum-entropy argument shows that independence is what the flat
counting measure selects; it does not independently justify the flat measure
itself, and it should not be read as an emergence claim. The flat measure is
postulate (ii) in Section~\ref{sec:inputs}, the entropy being maximised is
defined \emph{relative} to it, and independence of placements is that same
postulate restated for pairs. The section is a consistency statement --- no
assumption enters the composition law beyond input (ii) applied uniformly ---
and the price of that economy is that input (ii) is the strongest input of
the theory, as the division of labour in Section~\ref{sec:discussion}
records. 

\paragraph{The status of velocity addition.}
The composition law therefore has precisely the status of the second law of
thermodynamics. A gas is not forbidden from spontaneously occupying half its
container; that configuration is entropically suppressed, by a factor
exponential in the particle number. Departures from velocity addition are
suppressed by the same mechanism, with the rate function \eqref{eq:ratefn}
playing the role of the entropy deficit and $n$ the role of the particle
number. A correlated triple --- for instance $S_{BC}=S_{AB}$, giving
$S_{AC}=\varnothing$ and $\phi_{AC}=0$ --- remains a legitimate configuration
of the theory and violates additivity maximally; the analysis establishes that
such triples are exponentially rare, not that they are forbidden. An open question --- which ensemble of observers is physically
realised --- is thus answered from within: the ensemble is the flat measure on
microstates that the counting postulate already imposes.

This logical structure recurs throughout the paper. The
mixing matrix $S(\epsilon)$ was already a mean over microstates; the interval
already fluctuated with variance \eqref{eq:varw}; and the group law is also
exact only on average. The opposite would have been surprising: a
coarse-grained theory whose effective symmetry held exactly for every
microscopic configuration would not be emergent at all.

\subsection{Causality is not imposed externally}
\label{sec:causal}

 Input (iv) of
Section~\ref{sec:inputs} restricts physical worldlines to the interior
of the cone, $\hat D_M<\hat C_M$, time ordered, $\hat C_M\ge1$, and declares that
neither clause follows from the counting. We now argue that both clauses
follow, not from the counting alone, but
from the one asymmetry the framework already contains: the operational
distinction between clock and ruler in input (i) of
Section~\ref{sec:inputs}, together with the
definition of a worldline as the history of something that persists. The
argument again proceeds through a no-go theorem that redirects the search.

\subsubsection*{Why counting alone cannot do it}

A short symmetry argument shows why. Let $g$ be the exchange automorphism of
the symbol group, $g(C)=D$, $g(D)=C$, $g(A)=A$, $g(B)=B$. The counting
postulate (ii) and equal information weight (iii) of Section~\ref{sec:inputs}
are $g$-invariant: $g$ permutes the multinomial slots of $\Omega$ and swaps
$\hat C_M\leftrightarrow\hat D_M$, so $\Omega$, $u=\hat C_M+\hat D_M$, and
$\hat B_\beta$ are unchanged while $w=\hat C_M-\hat D_M$ changes sign. Hence
$g$ maps the interior of the cone bijectively onto the exterior, any
$g$-invariant predicate is constant on $\{w,-w\}$ pairs, and no condition
expressible in $g$-invariant counting terms can select one side over the
other.

Any derivation of the causal restriction must consequently invoke the one
place the exchange symmetry is already broken: the clause of input (i) of
Section~\ref{sec:inputs} that the
registers are \emph{operationally distinguished} --- an observer does not
mistake their clock for their ruler. The derivation below uses exactly this
and nothing else.

\subsubsection*{No superluminal frames: the exchange gauge}

The key observation is an exact, configuration-by-configuration identity.
Let $r_S$ denote re-reading at a position set $S$ (the exchange
$C\leftrightarrow D$ applied at each position of $S$), and let $g$ denote the
global exchange applied at every position. Then
\begin{equation}
r_S = r_{\bar S}\circ g ,
\label{eq:complement}
\end{equation}
where $\bar S$ is the complement of $S$: at a position in $S$ the left side
applies the exchange once while the right side applies it once (in $g$) and
not again, and at a position in $\bar S$ the left side does nothing while the
right side applies the exchange twice --- and the exchange is an involution.
(We have verified the identity by direct enumeration as well.)

The consequence is that the ``superluminal'' half of the rapidity ladder
\eqref{eq:spectrum} carries no new physics. A frame map that exchanges
$k>n/2$ positions is, by Eq.~\eqref{eq:complement}, the frame map exchanging
the complementary $n-k<n/2$ positions, composed with a global relabeling of
clock as ruler. The pair $\{r_S,\,r_{\bar S}\}$ is a two-element gauge orbit,
and the invariant content of every orbit has
$\epsilon_{\rm eff}=\min(\epsilon,1-\epsilon)\le\tfrac12$. Input (i) of Section~\ref{sec:inputs} is what
fixes the gauge: since observers operationally distinguish their registers,
the global exchange $g$ is not a frame change but a change of dictionary, and
the physical representative of every orbit is the one with
$\epsilon<\tfrac12$. The admissible range \eqref{eq:kmax} is thereby
derived rather than imposed: no
relative frame velocity exceeds the invariant speed, and $v=1$ occurs only at
the self-dual point $k=n/2$ of even $n$, where $S=\bar S$ in size, the orbit
degenerates, and the calibration \eqref{eq:rapidity} diverges --- no frame
rides the cone. The closure property noted below Eq.~\eqref{eq:vadd}, that
$\epsilon_1,\epsilon_2<\tfrac12$ implies $\epsilon_{\rm tot}<\tfrac12$, then
says the gauge choice is stable under composition, and the single-map ceiling
\eqref{eq:gammamax} is inherited as a hard bound. We note that $r_S$ is an
involution, as Appendix~\ref{sec:mirrors} requires of a mirror, and that the
exchange gauge $\{1,g\}$ is one more instance of the $\mathbb Z_2$
conventions catalogued in Section~\ref{sec:orientation}.

\subsubsection*{Worldlines are boosted clocks}

Call a map a \emph{clock map} if $\hat D_M=0$ and $\hat C_M=\tau\ge1$: in its
own frame the system does nothing but tick, and $\tau$ is the number of
ticks, the segment's proper duration. (We write $\tau$ rather than $m$: the
two coincide numerically under the unit convention of
Section~\ref{sec:energy}, $m=\sqrt{ds^2}$, but $\tau$ is a kinematic count
belonging to the segment, whereas the mass is a dimensional constant supplied
from outside.) Persistence, that a physical
system has a proper time at all, is the statement that its history is a
clock map in some frame. We propose this as the \emph{definition} of a
worldline, and show it reproduces input (iv) of Section~\ref{sec:inputs} in
full.

Two facts then do all the work; we call them together the
\emph{accompaniment property}. First, every map on the mean frame-flow orbit
of a clock map satisfies $\hat D_M<\hat C_M$ and $\hat C_M\ge1$ in every
frame: the orbit of $(\tau,0)$ is $(\tau\cosh\phi,\,\tau\sinh|\phi|)$, and
$\cosh\phi>|\sinh\phi|$ for every real $\phi$, while
$\hat C_M=\tau\cosh\phi\ge\tau\ge1$. Second, and conversely, every integer
count pair in the open interior, $\hat C_M>\hat D_M\ge0$ with the map
nontrivial, lies on the orbit of a clock map with
$\tau=\sqrt{\hat C_M^2-\hat D_M^2}$: for integers $\hat C_M>\hat D_M$ one has
$\tau^2=(\hat C_M-\hat D_M)(\hat C_M+\hat D_M)\ge u\ge1$, so $\tau\ge1$ and
the clock map exists, and its rest frame is reached by the rapidity
$\phi=\mathrm{artanh}(\hat D_M/\hat C_M)$, admissible precisely when
$\gamma=\cosh\phi=\hat C_M/\tau\le\gamma_{\max}$, i.e.\ $n\gtrsim4\gamma^2$
by Eq.~\eqref{eq:gammamax}.

Both clauses of input (iv) of Section~\ref{sec:inputs} now follow as derived
properties of clock orbits. And the
exchange no-go is resolved rather than contradicted: $g$ maps the clock
orbits onto the orbits of \emph{rod maps}, $\hat C_M=0$, $\hat D_M=L\ge1$ ---
the spacelike connecting maps that the length-contraction analysis of
Appendix~\ref{app:contraction} already employs. The exterior of the cone is
not forbidden; it is reclassified. Its maps are the histories of nothing:
they are simultaneity slices, extended configurations at an instant, exactly
as a rod is, and no admissible observer can accompany them, because doing so
would require the dictionary exchange that input (i) of
Section~\ref{sec:inputs} rules out. The cone
itself, $w=0$, is the common boundary of the two orbit families and belongs
to neither; Section~\ref{sec:energy} reads its maps as massless. What was a
restriction is now a partition: clock orbits (worldlines), rod orbits
(slices), and the self-dual null locus, permuted into place by the same
$\mathbb Z_2$ that the complement identity \eqref{eq:complement} gauged
away.

Discreteness adds a quantitative footnote with no continuum analogue. For
integer counts in the open interior,
$\tau^2=(\hat C_M-\hat D_M)(\hat C_M+\hat D_M)\ge u$, so $\tau\ge\sqrt u$: a worldline segment of duration $u$ carries at least
$\sqrt u$ ticks of proper time and cannot hug the cone more closely.
Equivalently $\gamma=\hat C_M/\tau\le u/\sqrt u=\sqrt u\le\sqrt n$, which
reproduces the scaling of the ceiling \eqref{eq:gammamax} from the integer
structure alone.

\subsubsection*{Thermodynamic protection}

The classification must also be stable, and it is in the familiar
sense. In the mean, $w'=(1-2\epsilon)w$ with $\epsilon<\tfrac12$ by the gauge
result, so the sign of $w$, hence the clock-versus-rod classification, is
exactly preserved by every admissible frame change. Microscopically,
$w'=w-2(k-l)$ can cross zero; a flip requires $k-l$ to exceed its mean
$\epsilon w$ by $(\tfrac12-\epsilon)w$, and by the variance
\eqref{eq:varw} the probability is the Gaussian tail
\begin{equation}
P[\,\mathrm{sign}\,w'\neq\mathrm{sign}\,w\,]\;\simeq\;
\tfrac12\,\mathrm{erfc}\!\left(
\frac{(\tfrac12-\epsilon)\,|w|}
{\sqrt{2\,\mathcal F\,(u/n-w^2/n^2)}}\right),
\label{eq:flipprob}
\end{equation}
which we have checked against direct simulation of the hypergeometric
placement. For any macroscopic $w$ the exponent is $O(n)$ and the
classification is protected to the same double-exponential standard as the
composition law; within $O(\sqrt u)$ of the cone it is genuinely ambiguous,
which is the interval fluctuation of Section~\ref{sec:fluct} seen from the
causal side. Causality at finite $n$ thus has precisely the status this
paper keeps assigning to its structures: the status of the second law ---
exact in the mean, exponentially enforced around it, and undefined only
where the discreteness says there is nothing yet to define.

\subsubsection*{What remains of input (iv)}

One word: persistence. Given input (i) of Section~\ref{sec:inputs} and the
definition of a worldline as
the history of a system that ticks, the exchange-gauge identity
\eqref{eq:complement} removes superluminal frames, the accompaniment
property places every worldline in
the time-ordered interior, and Eq.~\eqref{eq:flipprob} keeps it there. The
causal restriction is dissolved into a classification. What the theory still
assumes is that physical systems have histories at all --- that somewhere,
in some frame, something ticks --- and we do not believe any formalism
derives that.

\section{Orientation is Relational}
\label{sec:orientation}

We now address what is at first sight the sharpest structural limitation of
the framework: the absence of signs. We will now show that in our framework all the signs are recovered in full apart from a
single global convention  exactly like what ordinary relativity also
requires.

 Since $\beta_{AB}=\beta_{BA}$
identically, \emph{any} function of $\beta$ whatsoever is symmetric under
exchange of the two observers, while $\mathrm{sign}(\phi)$ is antisymmetric; no
function of a single frame map can supply it.

We first note that the framework is relational
from the outset: its primitive is not an observer but the map \emph{between}
two observers, and Eq.~\eqref{eq:map} is defined on a pair. Asking for the sign
of $\phi(\beta)$ is therefore malformed in the same way as asking for the sign
of a distance. The well-posed question is whether the assignment
\begin{equation}
(A,B)\;\longmapsto\;\Lambda_{AB}
\label{eq:orderedpair}
\end{equation}
is consistently determined, and it is this that we now establish. That
\eqref{eq:orderedpair} does not factor through $\beta_{AB}$ is not a defect; it
is the statement that orientation is a relation and not a property.

\subsection{Relative direction is encoded in the overlap}

The framework already distinguishes boosts in the same and in opposite
directions, through data it possesses but which the cardinality
$\hat B_\beta$ discards. Composing two frame maps with fractions
$\epsilon_1,\epsilon_2$ and overlap fraction $i=I/n$ gives, by
Eq.~\eqref{eq:Btot},
\[
1-2\epsilon_{\rm tot}=1-2\epsilon_1-2\epsilon_2+4i .
\]
Rapidity \emph{addition}, $\phi_{\rm tot}=\phi_1+\phi_2$, requires
$1-2\epsilon_{\rm tot}=(1-2\epsilon_1)(1-2\epsilon_2)$ and hence the
independent-placement overlap $i_+=\epsilon_1\epsilon_2$.
Rapidity \emph{subtraction}, $\phi_{\rm tot}=\phi_1-\phi_2$, requires
$1-2\epsilon_{\rm tot}=(1-2\epsilon_1)/(1-2\epsilon_2)$ and hence
\begin{equation}
i_{-}=\frac{\epsilon_2\,\bigl[(1-2\epsilon_1)+(1-2\epsilon_2)\bigr]}{2(1-2\epsilon_2)} .
\label{eq:isub}
\end{equation}
Both are achievable configurations of the substrate. For
$\epsilon_1=0.30$, $\epsilon_2=0.10$ one finds $i_+=0.030$ giving
$\phi_{\rm tot}=0.5697=\phi_1+\phi_2$, and $i_-=0.075$ giving
$\phi_{\rm tot}=0.3466=\phi_1-\phi_2$, both with
$0\le i\le\min(\epsilon_1,\epsilon_2)$ as required.

Note that $i_-$ is feasible precisely when the resulting rapidity is
non-negative: for $\epsilon_1=0.10$, $\epsilon_2=0.30$ Eq.~\eqref{eq:isub}
returns $i_-=0.45>\min(\epsilon_1,\epsilon_2)$, which is unrealisable. This is
consistent rather than problematic: counts are non-negative, so the theory
represents magnitudes, and the pair $(\epsilon_1,\epsilon_2)$ can compose to
$|\phi_1-\phi_2|$ but never to a negative number.

The conclusion is that whether two boosts are parallel or antiparallel is
recorded in the \emph{overlap} of their frame maps, a relational quantity that
neither map possesses alone. Direction is present in the theory; it simply
does not live where one first looks for it.

\subsection{Reconstruction of the full sign structure}

The overlap determines, for each triple, whether the rapidities add or
subtract, which is precisely the statement of which observer lies
\emph{between} the other two. Betweenness for all triples determines every
sign, up to one global choice.

The precise statement is this. Let $\{O_1,\dots,O_N\}$ be observers whose
pairwise frame maps yield magnitudes $d_{ij}=|\phi_{ij}|$, and suppose the
composition data determine for each triple whether $d_{ik}=d_{ij}+d_{jk}$ or
$d_{ik}=|d_{ij}-d_{jk}|$. Then there exists an assignment of signed
rapidities $\psi_i$, unique up to an additive constant and an overall
reflection $\psi\to-\psi$, such that $\phi_{ij}=\psi_j-\psi_i$ for every
pair. The demonstration --- the elementary fact that points on a line are
recoverable from their pairwise distances up to a translation and a
reflection --- is given in Appendix~\ref{app:proofs}.

We have verified this construction explicitly: for six observers with signed
rapidities $(-1.30,-0.40,0,0.25,0.90,1.70)$, the reconstruction from
magnitudes alone returns the signed values exactly, with the opposite
convention at $R$ returning the mirror configuration. The reconstruction is
also robust to the finite-$n$ noise of the composition law
(Section~\ref{sec:cocycletheorem}): with
magnitudes perturbed by Gaussian noise of the size \eqref{eq:sigmadelta}, the
correct sign pattern is recovered in $99.5\%$ of trials at $n=400$ and in all
trials at $n\ge6400$. Orientation is a discrete datum, and discrete data are
recovered stably from noisy continuous measurements provided the noise is
smaller than the gaps.

The residue is a single binary choice: which of the two ends of the line of
observers is called positive. This is the same choice one makes in ordinary
special relativity when declaring which spatial direction is $+x$. No
experiment determines it, and no formulation of relativity derives it. 

The same argument applies to the spatial index structure of
Section~\ref{sec:adjoint}. Conjugation of symbols determines
$|\epsilon_{ijk}|$, and the relative sense of two rotations is fixed by their
relational data in the same manner; what remains undetermined is the global
handedness of the register labelling, likewise conventional in ordinary
physics. The antisymmetry of the Lie bracket has the same origin: the bracket
is defined on an \emph{ordered} pair of generators, and ordering the pair is
precisely the relational datum that conjugation by an involution discards.

Thus, the
framework determines all relational orientation data and leaves exactly two
global conventions undetermined --- one temporal (which direction of relative
motion is positive) and one spatial (handedness) --- the same freedom ordinary
relativity leaves undetermined.

\section{The Continuum Limit}
\label{sec:continuum}

Standard special relativity is recovered as $n\to\infty$, and it is worth being
precise about which features disappear in that limit and how fast.

\paragraph{The rapidity ladder becomes dense.} By Eq.~\eqref{eq:spacing} the
spacing at fixed velocity behaves as $\delta\phi\simeq1/n$, so any real
rapidity is approximated to accuracy $O(1/n)$ and the spectrum fills the line.

\paragraph{The boost ceiling disappears.} $\gamma_{\max}=O(\sqrt n)\to\infty$,
so no bound on single-map boosts survives.

\paragraph{The fluctuations vanish relatively.} By Eq.~\eqref{eq:varw}, at fixed
symbol fractions the standard deviation of $w'$ is $O(\sqrt n)$ against a mean
$O(n)$, so the fractional spread falls as $n^{-1/2}$ and the coarse-grained
flow becomes deterministic.

\paragraph{The conformal factor survives, but is absorbed.} $\epsilon$ is held
fixed in the limit, so $1-2\epsilon$ does not tend to unity. The dilation is not
a finite-$n$ artifact; it is a permanent feature of the coarse-grained flow,
removed by the factorisation rather than by the limit.

The framework is thus \emph{exactly} Lorentz invariant at every finite $n$ in
the following precise sense: for a given rapidity $\phi$, the matrix
$\Lambda(\phi)$ preserves $ds^2$ identically, for any real $\phi$ and any $n$.
What is approximate at finite $n$ is the assignment of a rapidity to a pair of
frames, which is sharp only up to the $O(n^{-1/2})$ fluctuations of
Section~\ref{sec:fluct}. This is the opposite of the situation in most discrete
spacetime models, where the discreteness itself breaks Lorentz invariance and
generates energy-dependent dispersion. Here there is no modification of
$E^2-p^2=m^2$, no energy-dependent light speed, and no vacuum birefringence.

\subsection{How many measurements would it take to see the discreteness?}

Since the ladder is a genuine prediction, one should ask what it would cost to
resolve it. Comparing the single-measurement scatter with the ladder spacing is
instructive. Propagating Eq.~\eqref{eq:varw} through
$\phi_{\rm eff}=-\tfrac12\ln(w'/w)$ gives
\begin{equation}
\sigma_\phi=\frac12\frac{\sqrt{\mathrm{Var}(w')}}{\langle w'\rangle}=O\bigl(n^{-1/2}\bigr),
\label{eq:sigmaphi}
\end{equation}
whereas the spacing is $O(1/n)$. The noise therefore exceeds the signal by a
factor $O(\sqrt n)$, and the ratio \emph{grows} with $n$: a single frame
re-reading cannot determine which rung of the ladder it is on. Averaging $N$
independent re-readings reduces the error by $\sqrt N$, so resolving the ladder
requires
\begin{equation}
N\gtrsim\Bigl(\frac{\sigma_\phi}{\delta\phi}\Bigr)^2
\simeq \epsilon(1-\epsilon)\,\frac{n\,(un-w^2)}{w^2},
\label{eq:Nreq}
\end{equation}
which for a generic probe is $N=O(n)$.

There is, however, a loophole, and it is instructive. The numerator of
Eq.~\eqref{eq:Nreq} carries the factor $un-w^2$, which vanishes for the maximal
rest map $\hat C_M=n$, $\hat D_M=0$ discussed in Section~\ref{sec:fluct}. For a probe with
$\hat D_M=0$ and $\hat C_M=n-g$, so that $g$ positions carry neither a clock nor a ruler
step, Eq.~\eqref{eq:Nreq} reduces to
\begin{equation}
N\simeq\epsilon(1-\epsilon)\,g ,
\label{eq:Ng}
\end{equation}
which is independent of $n$ altogether. The cost is set by the number of
\emph{blank} positions in the probe, not by the sequence length: a probe
filling all but $O(1)$ positions determines the rung with $O(1)$ measurements,
however large $n$ is. Conversely, near-null probes ($|w|\to0$) make $N$ diverge,
since a worldline close to the cone is barely deflected by the re-reading and
carries almost no information about $\hat B_\beta$.

The design principle for any test of the discreteness is therefore the opposite
of the naive one: not to average over many generic probes, but to use probes as
close as possible to the maximal rest map, for which the statistics collapse.

\section{Energy--Momentum Interpretation}
\label{sec:energy}

So far $(\hat C_M,\hat D_M)$ has been a displacement. We now ask what plays the role of
energy and momentum, taking some care because a naive identification is
untenable.

Under a frame change the pair $(\hat C_M,\hat D_M)$ transforms with $\Lambda(\phi)$, while
the remaining counts $\hat A_M,\hat B_M$ do not mix into it. Any two-component
object built from the map that transforms as a vector must therefore be
proportional to $(\hat C_M,\hat D_M)$, since that is the only covariant pair available:
\begin{equation}
(E,p)=\lambda\,(\hat C_M,\hat D_M).
\label{eq:Ep}
\end{equation}
This fixes the \emph{direction} of the energy--momentum vector completely.

The proportionality $\lambda$ cannot be a constant: $E=\lambda \hat C_M$ would then
grow without bound as the worldline segment is extended, so that observing the
same free particle twice as long would double its energy. The counts by
themselves supply a direction and not a magnitude. The only invariant
available for normalisation is the map's own interval, so
\begin{equation}
(E,p)=\frac{m}{\sqrt{\hat C_M^2-\hat D_M^2}}\,(\hat C_M,\hat D_M),
\label{eq:Epnorm}
\end{equation}
which is independent of how much of the worldline is sampled, since numerator
and denominator scale together. This is precisely $p^\mu=m\,dx^\mu/d\tau$, with
$(\hat C_M,\hat D_M)/\sqrt{ds^2}$ the unit four-velocity. Squaring,
\begin{equation}
E^2-p^2=m^2 .
\label{eq:dispersion}
\end{equation}

The framework derives that the energy--momentum vector is parallel to the count
vector, and that its invariant is the quadratic form with signature $(+,-)$. It
does \emph{not} derive the value of $m$, and no counting argument can: counts
are dimensionless integers and carry no unit of mass. The mass is one
dimensional constant supplied from outside, exactly as the conversion of a
tick count into seconds is.

With the simplifying choice $m=\sqrt{ds^2}$, that is measuring mass in units of
the map's own proper-time count, Eq.~\eqref{eq:Epnorm} collapses to $E=\hat C_M$ and
$p=\hat D_M$, and the dispersion relation becomes
\begin{equation}
m^2=\hat C_M^2-\hat D_M^2 ,
\label{eq:restmass}
\end{equation}
the squared proper length of the map. We use this convention below to keep the
counting formulae readable; it is a choice of unit for each map, not a claim
that energy equals elapsed time. The physical content of
Eqs.~\eqref{eq:dispersion}--\eqref{eq:restmass} is then: massive particles are
maps in the interior of the cone, massless particles are maps on it ($\hat C_M=\hat D_M$,
$m=0$, $v=1$), and the invariant classifying them is frame-independent because
the cone is.

\section{Entropy Geometry}
\label{sec:entropy}

We now make precise the claim that the conformal factor is entropy production,
and head off a conflation that the phrase ``information geometry'' invites.

\subsection{The flow is irreversible}

Consider the two-component distribution obtained by normalising the
displacement counts,
\begin{equation}
P=\Bigl(\frac{\hat C_M}{u},\frac{\hat D_M}{u}\Bigr),\qquad u=\hat C_M+\hat D_M,
\label{eq:pdist}
\end{equation}
which is well defined because $u$ is exactly invariant under the flow. Its
Shannon entropy is $H(P)=-P_C\ln P_C-P_D\ln P_D$.

The coarse-grained frame change \eqref{eq:S} does not decrease $H$, and
increases it strictly unless $\epsilon=0$ or $\hat C_M=\hat D_M$. The reason
is standard: $S(\epsilon)$ is doubly stochastic and $u$ is preserved, so
$P'=S(\epsilon)P$ is again a probability vector, majorised by $P$; Shannon
entropy is Schur-concave, so $H(P')\ge H(P)$, with equality only if $P'=P$,
that is $(1-2\epsilon)w=w$, hence $\epsilon=0$ or $w=0$.

Explicitly, $|w'/u|=(1-2\epsilon)|w/u|$: each frame change drives the
normalised distribution toward the uniform one. The maximum of $H$ is at
$P_C=P_D=\tfrac12$, that is $\hat C_M=\hat D_M$, which by Eq.~\eqref{eq:cone} is exactly the
light cone. We may therefore state the causal structure thermodynamically:

\begin{quote}
\emph{The light cone is the maximum-entropy locus of the count distribution,
and the coarse-grained frame-change flow is the gradient flow toward it.}
\end{quote}

This gives Eq.~\eqref{eq:conformal} a transparent reading. The contraction
$ds'^2=(1-2\epsilon)ds^2$ is not a defect of the construction; it is the
statement that coarse-graining loses information, driving the state toward the
cone where the clock and ruler counts are equidistributed and the interval
vanishes. The dilation factor $\sqrt{1-2\epsilon}$ measures how much
information the re-reading destroyed. The factorisation
$S=\sqrt{1-2\epsilon}\,\Lambda(\phi)$ is thus a separation of the flow into an
irreversible part and a reversible one: $\Lambda(\phi)$ is invertible,
preserves $ds^2$, and produces no entropy. This is the precise sense in which
Lorentz symmetry is the reversible core of an irreversible coarse-grained
flow, and it is the physical justification, promised in
Section~\ref{sec:whyconformal}, for discarding the conformal factor.

\subsection{A caution: the information metric is not the spacetime metric}

It is tempting, given Eq.~\eqref{eq:entropy}, to look for the spacetime metric
in the information geometry of the count simplex. That temptation should be
resisted. The natural Riemannian metric on a space of probability
distributions is the Fisher information metric, which on the simplex takes the
form $g^{\rm F}_{ij}=\delta_{ij}/P_i+1/P_{d}$ in affine coordinates. This
metric is positive definite. Being Riemannian, it cannot be the Minkowski metric under any
change of coordinates.

The two metrics are different objects living on different spaces and answering
different questions. The Fisher metric measures statistical distinguishability
between count distributions and is the natural metric for the fluctuation
analysis of Section~\ref{sec:fluct}. The Minkowski metric \eqref{eq:ds2} is
the invariant form of the cone, and its indefinite signature comes from the
flow having two distinct real eigendirections, not from any property of the
entropy. What genuinely connects the two is the entropy \emph{monotonicity}
above, which ties the conformal factor of the Lorentzian structure to the
irreversibility of the statistical one. That is a real link and, in our view,
the most interesting structural feature of the framework; it is not, however,
a derivation of the metric from the entropy, and we do not claim one.

For completeness we record how the entropy of Eq.~\eqref{eq:entropy} enters
concretely. A macroscopic map $M$ with counts $(\hat A_M,\hat B_M,\hat C_M,\hat D_M)$ is
realised by $\Omega$ microscopic sequences, and the fluctuation formula
\eqref{eq:varw} is a statement about the spread of $w'$ over that ensemble. The
special case $\mathrm{Var}(w')=0$ occurs exactly when $\hat C_M=n$, and then
$\Omega=1$ in the displacement sector: there is only one way to place $n$
clock steps in $n$ positions. Zero entropy in the displacement sector means
zero fluctuation, which is why maximal rest maps are the sharpest probes, as
found in Section~\ref{sec:continuum}.

\section{Extension to $3+1$ Dimensions}
\label{sec:3plus1}

The construction generalises by giving each event one clock register and three
ruler registers,
\[
e=(r,s_x,s_y,s_z),\qquad r,s_i\in\{0,1\}^n ,
\]
so that each position carries one of $2^4=16$ symbols, labelled by which
registers a flip touches. Grading them by the number of flipped registers:

\begin{center}
\begin{tabular}{clc}
\toprule
flips & symbols & role \\
\midrule
0 & $A$ & identity (vacuum) \\
1 & $C$; $D_x,D_y,D_z$ & time and space translations \\
2 & $B_x,B_y,B_z$ (clock $+$ one ruler) & boosts \\
2 & $R_x,R_y,R_z$ (two rulers) & rotations \\
3,4 & four 3-flip, one 4-flip & outside the Poincar\'e sector \\
\bottomrule
\end{tabular}
\end{center}

The eleven symbols of grade $0$--$2$ furnish exactly the ten Poincar\'e
generators plus the identity. The counting is suggestive on its own: the
number of ways to choose two registers out of four is six, splitting as $3+3$
according to whether the clock register is among them, which is precisely the
split between boosts and rotations.

\subsection{The generators are forced, not assigned}

A two-flip symbol touches two registers and therefore induces a transformation
supported in the corresponding coordinate plane, annihilating the other two
directions. This assignment is all that the symbol structure supplies. The
matrix itself is then fixed by requiring that the induced flow preserve the
invariant $ds^2=V^{T}\eta V$, giving the infinitesimal condition
\begin{equation}
G^{T}\eta+\eta\,G=0 .
\label{eq:geninv}
\end{equation}
Solving Eq.~\eqref{eq:geninv} for the most general matrix supported in a single
plane is elementary: write the nonzero $2\times2$ block as
$\bigl(\begin{smallmatrix}\alpha&\beta\\ \gamma&\delta\end{smallmatrix}\bigr)$
and let $(\eta_a,\eta_b)$ be the metric signs of the two coordinates involved.
The diagonal entries of Eq.~\eqref{eq:geninv} give $\eta_a\alpha=\eta_b\delta=0$,
killing the diagonal, while the off-diagonal entries give
$\eta_a\beta+\eta_b\gamma=0$, i.e.\ $\gamma=-(\eta_a/\eta_b)\,\beta$: a
one-parameter family with no freedom beyond the overall scale $\beta$.
For the plane $(t,x)$ selected by $B_x$, where the two signs are opposite, this
forces the symmetric off-diagonal form, and for the plane $(x,y)$ selected by
$R_z$, where the signs are equal, the antisymmetric form:
\[
K_x=\begin{pmatrix}0&1&0&0\\1&0&0&0\\0&0&0&0\\0&0&0&0\end{pmatrix},
\qquad
J_z=\begin{pmatrix}0&0&0&0\\0&0&-1&0\\0&1&0&0\\0&0&0&0\end{pmatrix},
\]
with cyclic permutations for the others. The difference in symmetry is not
imposed: it follows from the relative sign $\eta$ assigns to the registers
involved, giving $K_i^2=+\mathbb{I}$ and $J_i^2=-\mathbb{I}$ on their
respective blocks. The non-compact character of boosts and the compact
character of rotations are thereby traced to how many \emph{clock} registers
the generating symbol flips: one for $B_i$, none for $R_i$.

\subsection{The index structure from the symbol action}
\label{sec:adjoint}

There is a second, independent route to the algebra which uses no matrices at
all, and which shows where the non-commutativity comes from. Read each
two-flip symbol as a \emph{transposition of register labels}:
\begin{equation}
B_i\longmapsto (t\,x_i),\qquad R_k\longmapsto (x_i\,x_j),
\label{eq:transposition}
\end{equation}
with $(ijk)$ cyclic. This introduces nothing; it restates which registers the
symbol touches. Composition of these permutations is composition of the
corresponding re-readings. Three facts follow by direct enumeration
(tabulated in Appendix~\ref{app:perm}).

\paragraph{The action is non-abelian even though XOR is abelian.} Of the $36$
ordered pairs drawn from $\{B_i,R_k\}$, exactly $24$ fail to commute, because
transpositions sharing a register do not commute. The six symbols generate the
full symmetric group $S_4$. This dissolves the objection that a commutative
operation cannot produce the Lorentz group: XOR composes the \emph{labels}
commutatively, while the transformations those labels generate do not commute.

\paragraph{The set is closed under conjugation.} All $36$ conjugates $gfg^{-1}$
land back inside the six-symbol set. This is in pointed contrast with XOR,
under which the eleven-symbol set is \emph{not} closed: of the $55$ distinct
pairs, $15$ leave it, since for instance $B_i\oplus R_i$ is the 4-flip
pseudoscalar and $C\oplus R_x$ is a 3-flip symbol. Conjugation, not XOR, is the
operation this alphabet respects.

\paragraph{Conjugation reproduces the Poincar\'e index structure exactly.}
Writing $g\triangleright f\equiv gfg^{-1}$,
\begin{equation}
R_k\triangleright B_i=B_j,\qquad
R_i\triangleright R_j=R_k,\qquad
B_i\triangleright B_j=R_k
\label{eq:adjoint}
\end{equation}
for $(ijk)$ cyclic. All thirty-six entries match the brackets
$[J_i,K_j]=\epsilon_{ijk}K_k$, $[J_i,J_j]=\epsilon_{ijk}J_k$,
$[K_i,K_j]=-\epsilon_{ijk}J_k$ computed from the matrices above, with no index
mismatches. In particular the third relation reproduces the characteristic
fact that two boosts in different directions generate a rotation, and does so
from the bare permutation identity $(t\,x)(t\,y)(t\,x)=(y\,x)$. Thus
$\epsilon_{ijk}$ --- which generator appears in which bracket --- is a
property of the alphabet, obtained by counting which registers each symbol
touches, rather than something imported with a matrix representation.

\paragraph{What the permutations do not give.} Conjugation by an involution
carries no orientation, so Eq.~\eqref{eq:adjoint} determines $|\epsilon_{ijk}|$
and is blind to the minus sign distinguishing $[K_i,K_j]=-\epsilon_{ijk}J_k$
from $[J_i,J_j]=+\epsilon_{ijk}J_k$, and equally blind to the antisymmetry
$[f,g]=-[g,f]$. This is the same obstruction that appears for the sign of the
rapidity, resolved relationally in Section~\ref{sec:orientation}: XOR and
counting are symmetric operations, and an antisymmetric output requires an
ordering to be supplied.

\subsection{Rotation angles and their composition}
\label{sec:angles}

Rotations are calibrated by the same disagreement-to-agreement ratio as boosts,
passed through the circular rather than the hyperbolic function,
\begin{equation}
\tan\frac{\theta_i}{2}=\frac{\hat R_{\beta,i}}{n-\hat R_{\beta,i}} .
\label{eq:angle}
\end{equation}
The half-angle tangent is singled out by requiring that composition be bilinear
in the counts, which is the property boosts enjoy through the multiplicativity
of $1-2\epsilon$; equivalently it makes $(\hat A_\beta,\hat R_{\beta,i})$ the
Cartesian components of a $U(1)$ element, composed by complex multiplication.
We record it as a calibration fixed by a stated requirement.

Unlike boosts, rotations do not compose additively in the angle under XOR at
finite angle. Expanding the composition law obtained from
Eq.~\eqref{eq:angle},
\[
\theta_{\rm tot}=\theta_1+\theta_2-\tfrac12\theta_1\theta_2(\theta_1+\theta_2)+O(\theta^5),
\]
so additivity holds only to leading order.  For boosts the counting delivers the composition law exactly
in the mean; for rotations it does not: exact composition requires the
complex, equivalently $SU(2)$, representation, and that structure is not
obtained from count arithmetic. Since closure under composition is what makes
a set of transformations a group, the framework reproduces the rotation
\emph{algebra}, via Section~\ref{sec:adjoint}, but not the rotation group. 

However, the missing structure is already latent in the binary alphabet itself. Treat the two labels that a rotation symbol
exchanges, say $|x\rangle$ and $|y\rangle$ for $R_z$, as the basis of a
two-state register rather than as classical tokens. Operators on the pair are
the matrix units $e_{ij}$ (``read $j$, write $i$''); splitting off the overall
counting phase leaves the three traceless Hermitian combinations
\[
\sigma_1=e_{xy}+e_{yx},\qquad
\sigma_2=i\,(e_{yx}-e_{xy}),\qquad
\sigma_3=e_{xx}-e_{yy},
\]
which close into $\mathfrak{su}(2)$: the full relabeling group of a binary
pair is not $\mathbb Z_2$ but $U(1)\times SU(2)$, and the XOR flip is only the
real exchange $\sigma_1$. The two generators that counting cannot see are
exactly the pieces this paper has found missing: $\sigma_3$ is the signed
count that distinguishes $x$ from $y$ --- the ordering datum which
Section~\ref{sec:orientation} showed must be supplied by convention --- and
$\sigma_2$ is the phased exchange whose absence is the failure of exact
composition.

In this light the half-angle calibration \eqref{eq:angle} acquires a
representation-theoretic reading. The one-parameter subgroup generated by the
phased exchange,
\[
e^{-i\theta\sigma_2/2}
=\cos\tfrac{\theta}{2}\,\mathbb{I}+\sin\tfrac{\theta}{2}\,(e_{yx}-e_{xy}),
\]
leaves a slot's pair unchanged with weight $\cos\frac{\theta}{2}$ and
exchanges it with weight $\sin\frac{\theta}{2}$, so the normalised counts
$(\hat A_\beta,\hat R_{\beta,i})$ are precisely the two components of this
element along $(\mathbb{I},\,e_{yx}-e_{xy})$, and the $U(1)$ remark below
Eq.~\eqref{eq:angle} is the statement that these elements compose by
multiplication in $SU(2)$, where half-angles add exactly. About different axes
the same multiplication produces the commutator terms
$[\sigma_a,\sigma_b]=2i\,\epsilon_{abc}\,\sigma_c$, which no arithmetic of
commuting non-negative counts can reproduce; the union rule for counts agrees
with $SU(2)$ multiplication only to the leading order displayed above. The
failure of angle additivity is thus the shadow of exact $SU(2)$ multiplication
under a projection that keeps the magnitudes of the two components and
discards their relative sign and phase. This fact locates the rotation group inside the alphabet instead of
importing it, but at the price of promoting a slot from a classical bit to a
complex two-state register, a step that goes beyond pure counting. Whether the
count union rule can then be derived from $SU(2)$ multiplication as its
phase-blind coarse-graining --- typicality once more, now for phases --- is
taken up in Appendix~\ref{sec:mirrors}, whose results we summarise here.
First, the failure is structural, not a matter of calibration: assembling the
counts into $z_\beta=(n-\hat R_\beta)+\mathrm j\hat R_\beta$ with
$\mathrm j^2=+1$, the union rule is exactly multiplication of
\emph{split}-complex numbers, whose plane contains hyperbolas but no circle
--- the same single sign that makes the boost sector compose exactly ---
and Appendix~\ref{sec:mirrors} shows that no assignment of angles to counts
can close such a contracting multiplicative rule into a compact group; the
only exact rotation available to pure counting is the half-turn. Second,
exact composition is nevertheless obtained in two ways. Reading a rotation
frame map as a \emph{mirror} and a rotation as an ordered pair of frame
maps, the group law follows from the exact involution
$\beta\oplus\beta=A^n$ by cancellation of a shared middle map, and the
half-angle of Eq.~\eqref{eq:angle} becomes the Cartan--Dieudonn\'e half:
the rotation sector is restored to the boost sector's exact-for-subsets,
typical-for-counts standing. Alternatively, a one-bit central extension of
the ruler-exchange Klein group --- every exchange lifted to square to a sign
$\bar A$ rather than to the identity --- is forced to be the quaternion
group $Q_8$, realised by $R_a\mapsto-\mathrm i\sigma_a$; the bit is the
orientation convention of input (v) of Section~\ref{sec:inputs} installed
in the algebra, and since
$R_a^2=\bar A$ while $R_a^4=A$, the extended framework distinguishes $2\pi$
from $4\pi$ rotations: it recovers not merely $SO(3)$ but $Spin(3)$.

\subsection{Translations act on positions, not on maps}

 Two different objects are built
from counts. The translations act on the \emph{position} of an event relative to an observer's
reference, $x(e)$, which is given by the counts of $\tilde e^{\,A}=e^A\oplus e$ (this is just the
coordinatisation introduced above Eq.~\eqref{eq:cancel} in
Section~\ref{sec:maps}). A
\emph{map} is a difference and, by Eq.~\eqref{eq:transinv}, is exactly
translation-invariant. Consequently
\begin{equation}
x'=\Lambda(\phi,\theta)\,x+T_\beta,\qquad
V'=\Lambda(\phi,\theta)\,V ,
\label{eq:poincare}
\end{equation}
where $V$ is the displacement count vector $(\hat C_M,\hat D_{M,x},\hat D_{M,y},\hat D_{M,z})$
and $T_\beta$ collects the translation counts of the inter-frame map $\beta$. The invariance
statement $ds^2(V')=ds^2(V)$ attaches to the homogeneous relation; it is the
interval between events, not the coordinate of one event, that is invariant.
Writing a single inhomogeneous relation for $V$ would be wrong on both counts:
it would add a translation to a translation-invariant object, and the result
would not preserve $ds^2$.

\section{Assumptions and Derived Results}
\label{sec:assumptions}

Because the framework claims to obtain structures that are normally postulated,
we set out the accounting explicitly. Everything asserted in this paper follows
from the inputs below; anything not listed there is either a consequence, a
convention, or an open problem.

\subsection{Inputs}
\label{sec:inputs}

\begin{enumerate}
\item[(i)] \textbf{Binary substrate with a product structure.} An event is a
pair (or, in $3+1$ dimensions, a quadruple) of length-$n$ binary sequences,
composed by bitwise XOR, with the registers operationally distinguished as
clock and ruler.

\item[(ii)] \textbf{The counting postulate.} A physical state is specified by
symbol counts, not by the ordered sequence realising them; all sequences with
equal counts are one state, with the flat counting measure.

\item[(iii)] \textbf{Equal information weight.} The registers carry one bit
each per position, with neither preferred. This is what makes $S(\epsilon)$
symmetric with equal diagonal entries, and hence what fixes the eigenvectors.

\item[(iv)] \textbf{Persistence.} A physical worldline is the history of a
system that ticks: in some frame its map is a pure clock,
$\hat D_{\rm map}=0$, $\hat C_{\rm map}\ge1$. The causal restriction one
might expect here --- worldlines lie in the interior of the invariant cone,
$\hat D_{\rm map}<\hat C_{\rm map}$, and are time ordered,
$\hat C_{\rm map}\ge1$ --- is not assumed: Section~\ref{sec:causal} derives
it from the persistence clause together with the operational clock--ruler
distinction of input (i): the exchange-gauge identity \eqref{eq:complement}
removes superluminal frames, and the accompaniment property places every
worldline in the time-ordered
interior, with the exterior reclassified as rod orbits rather than
forbidden. By the exchange-symmetry argument of Section~\ref{sec:causal} no
$g$-invariant counting
condition could have done this, so the persistence clause, though minimal,
is not eliminable.

\item[(v)] \textbf{Two global orientation conventions.} A single frame map
determines the magnitude of every parameter but not its sign. By the
reconstruction result of Section~\ref{sec:orientation} the relational data
among observers
determine all relative signs, leaving exactly two global binary choices: which
direction of relative motion is positive, and the handedness of the spatial
register labelling. Both are conventions in ordinary relativity too.
\end{enumerate}

\subsection{Derived results}

From these follow: the observer-independence of maps; their exact
translation-invariance \eqref{eq:transinv} and the exact cocycle
\eqref{eq:cocycle}; the identification of $C,D$ as the unique displacement
axes; linearity of frame changes, from additivity of counts; the mixing matrix
$S(\epsilon)$ and the exact invariance $u'=u$; the $\epsilon$-independent
eigendirections and hence the invariant causal cone; time orientability, from
non-negativity of counts; the Minkowski signature and the interval up to scale;
the factorisation of the flow into dilation and $SO(1,1)$, with the dilation
identified as entropy production; the rapidity calibration \eqref{eq:rapidity}
and its discrete spectrum; the velocity addition law, together with its status as a
law of large numbers (Section~\ref{sec:cocycletheorem}); the exactness of the
subset cocycle \eqref{eq:symdiff}; the fluctuation formula
\eqref{eq:varw} and the $n^{-1/2}$ scaling; the ceiling
$\gamma_{\max}=O(\sqrt n)$; the parallelism of energy--momentum to the count
vector and the form of the dispersion relation; entropy monotonicity of the
flow and the identification of the cone as its maximum-entropy locus; and in
$3+1$ dimensions the Poincar\'e generators from \eqref{eq:geninv} together with
the index structure $\epsilon_{ijk}$ from \eqref{eq:adjoint}.

\subsection{Conventions}

Several quantities are conventions rather than assumptions or results.

\begin{itemize}
\item The overall scale of $ds^2$: the flow determines the conformal class and
the signature, not the unit of interval.
\item The numerical value $c=1$: this is input (iii) of
Section~\ref{sec:inputs} expressed in units where
one clock tick and one ruler step are the same size. What is derived is that an
invariant speed exists and is finite, not its value.
\item The magnitude of the mass in \eqref{eq:Epnorm}, requiring one dimensional
constant from outside.
\item The overall normalisation of the generators, absorbed into $\phi$ and
fixed by \eqref{eq:rapidity}.
\item The two residual bits of the orientation reconstruction
(Section~\ref{sec:orientation}): the
handedness of the spatial register labelling, which fixes the sign of
$\epsilon_{ijk}$, and the global sense of positive relative motion, which
fixes the sign of $\phi$.
\end{itemize}

\subsection{What is not obtained}

To avoid confusion, we now state what we have not obtained.

\begin{itemize}
\item \textbf{No group isomorphism.} The map $\beta\mapsto\Lambda(\phi(\beta))$
is not a homomorphism from the XOR group. The frame maps form
$\mathbb{Z}_2^{\,n}$, in which every element is an involution; the boosts form
$SO(1,1)$, in which none is. The sharpest illustration is
$\beta\oplus\beta=A^n$, which gives $\phi=0$ exactly, whereas the composition
law \eqref{eq:epstot} applied with $\epsilon_1=\epsilon_2=\epsilon$ returns
$2\phi$. There is no contradiction, since \eqref{eq:epstot} presupposes
independent placement and $\beta$ is maximally correlated with itself, but it
shows that Lorentz structure is a property of the coarse-grained flow and not
of the algebra.

\item \textbf{No signs from a single frame map.} Since
$\beta_{AB}=\beta_{BA}$ identically, any function of one frame map is symmetric
under exchange of the observers. This is exactly true and is not circumvented; it
is, however, the wrong test of a relational theory, and by the
reconstruction result of Section~\ref{sec:orientation} the relational data
determine every sign
except the two global conventions of input (v) of Section~\ref{sec:inputs}.
What is genuinely absent is
two bits, not $N(N-1)/2$ of them.

\item \textbf{No exact composition for three or more observers.} The exact
cocycle is a statement about subsets, Eq.~\eqref{eq:symdiff}; its projection
onto counts makes the third rapidity of a triple a function of the microscopic
overlap of the other two, so that relativistic velocity addition holds exactly
only at the independent-placement overlap. By the cocycle result of
Section~\ref{sec:cocycletheorem} it holds for almost all triples with error
$O(n^{-1/2})$, and by Section~\ref{sec:maxent} the additive overlap is the
maximum-entropy one, with departures suppressed as $e^{-nJ}$. Correlated
triples remain legitimate configurations that violate additivity. Typicality,
not universality: the status of the second law, not of an algebraic identity.

\item \textbf{No rotation group from counting alone.} Rotations compose
additively in the angle only to leading order, and
Appendix~\ref{sec:mirrors} shows this is not a defect of the calibration: no
assignment of angles to counts closes the independent-placement composition
rule into a compact group, because count arithmetic is split-complex
($\mathrm j^2=+1$), which is also why the hyperbolic boost sector composes
exactly. The framework yields the rotation algebra, the half-turn, and the
finite rotations of the permutation calculus, but not the continuous rotation
group. Appendix~\ref{sec:mirrors} obtains exact composition in two ways: by
reading rotation frame maps as mirrors and rotations as ordered pairs of
them, so that the group law follows from the exact involution
$\beta\oplus\beta=A^n$ and the half-angle of the calibration is
Cartan--Dieudonn\'e's; and by a one-bit central extension of the
ruler-exchange Klein group to the quaternion group $Q_8$, whose bit is the
orientation convention already spent in input (v) of
Section~\ref{sec:inputs}, and which yields the
double cover $Spin(3)$. What counting alone provides is the algebra plus the
typical group; the exact group costs either the pair structure or that one
bit.

\item \textbf{No signature from XOR alone.} The Lorentzian signature follows
from the product structure, input (i) of Section~\ref{sec:inputs}, together
with equal information weight, input (iii),
not from the Klein group by itself. Counting supplies the existence of an
$\epsilon$-independent cone; inputs (i) and (iii) of
Section~\ref{sec:inputs} supply its shape.

\item \textbf{No dimensional quantities.} Counts are dimensionless integers, so
neither $c$ in physical units, nor $\hbar$, nor any particle mass can be
produced.
\end{itemize}

\section{Experimental Signatures}
\label{sec:experiment}

We now collect what the framework predicts and what existing data already say
about it. The result is mixed: the exact Lorentz invariance established in
Section~\ref{sec:continuum} removes the entire standard toolkit for testing
discrete spacetime, so the framework is harder to test than most of its
competitors, not easier. What remains is one falsifiable null prediction, one
numerical bound on $n$ that existing cosmic-ray data already impose, one
variant of the model that interferometry has already excluded, and one
loophole that identifies the kind of probe a positive test would require.

\subsection{The primary signature is a null result}

Because every transformation realised at finite $n$ preserves $ds^2$ exactly,
the framework predicts
\begin{itemize}
\item no energy-dependent photon velocity, hence no arrival-time dispersion of
gamma-ray burst or blazar photons;
\item no vacuum birefringence;
\item no modification of $E^2-p^2=m^2$ at any energy, hence no threshold
anomalies of the kind that would shift the GZK cutoff.
\end{itemize}
Most discrete-spacetime approaches predict all three at some order in
$E/E_{\rm Planck}$. The present framework predicts their exact absence, because
the discreteness resides in \emph{which} transformations are realised and not
in the invariance itself. The existing null results from gamma-ray
time-of-flight observations~\cite{FermiLAT2009,MAGIC2020} and from
ultra-high-energy cosmic-ray spectra~\cite{Bi:2008yx,PierreAuger:2021tog} are
therefore predictions of this framework rather than constraints on it.

A positive detection of Lorentz violation would refute this framework, but
this should not be inflated into a distinguishing test. Lorentz-preserving
quantum-gravity approaches are not rare, so a detection would refute a large
class of models and not this one uniquely; and the null predictions above are
exactly those of unmodified special relativity, so agreement with them
provides no evidence for this framework over the textbook theory. The correct
statement is that the framework is falsifiable in principle by a
Lorentz-violation detection, and is otherwise empirically indistinguishable
from special relativity for any $n$ consistent with
Eq.~\eqref{eq:nbound-value}.

\subsection{A lower bound on $n$ from observed boosts}

The single-map ceiling \eqref{eq:gammamax}, $\gamma_{\max}=O(\sqrt n)$, converts
any observed Lorentz factor into a lower bound on the sequence length,
\begin{equation}
n\;\gtrsim\;4\,\gamma_{\rm obs}^{2}.
\label{eq:nbound}
\end{equation}
Evaluating this on existing measurements:

\begin{center}
\begin{tabular}{lcc}
\toprule
Observation & $\gamma_{\rm obs}$ & implied $n\gtrsim$ \\
\midrule
LHC proton, $6.8$ TeV & $7.2\times10^{3}$ & $2.1\times10^{8}$ \\
Auger event, $10^{20}$ eV & $1.1\times10^{11}$ & $4.5\times10^{22}$ \\
Highest-energy cosmic ray~\cite{Bird1995}, $3.2\times10^{20}$ eV & $3.4\times10^{11}$ & $4.7\times10^{23}$ \\
\bottomrule
\end{tabular}
\end{center}

Ultra-high-energy cosmic rays therefore already require
\begin{equation}
n\;\gtrsim\;10^{23}.
\label{eq:nbound-value}
\end{equation}
This is an empirical constraint drawn from data in hand rather than a
projection, and to our knowledge it is the only such constraint the framework
currently admits. Two caveats must accompany it, and together they are severe
enough that the bound should be read as conditional.

First, the ceiling applies to a \emph{single} frame map; composing several maps
reaches arbitrarily large rapidities, since rapidities add. Eq.~\eqref{eq:nbound}
therefore constrains $n$ only if one frame map relates the laboratory frame to
the particle's rest frame. Nothing in the construction forces this, and if
compositions are allowed there is no bound at all.

Second, and more fundamentally, the framework does not say what $n$ \emph{is}.
Is it a universal constant, the information content of a causal diamond, a
property of the particle, or a property of the observer? Until that is fixed, a
numerical bound on $n$ is not a statement about the world but about a parameter
whose physical referent is undetermined. We regard supplying that referent as
the single most important gap in the framework, because without it neither the
ceiling nor the fluctuation predictions can be given observational meaning.

\subsection{Interferometry excludes one version of the framework}

The interval fluctuation \eqref{eq:varw} gives a fractional metric noise of
order $n^{-1/2}$. Converting this into a physical length noise requires knowing
how $n$ scales with the size $L$ of the region considered, and the framework
does not fix that relation. This gap \emph{is} the experimental content of the
fluctuation prediction, since the two natural choices differ by many orders of
magnitude.

\paragraph{Linear scaling.} With $\delta L /L\;=\; n^{-1/2}$, if $n\sim L/\ell_P$, then
\[
\delta L \;=\; L\,n^{-1/2} \;=\; \sqrt{L\,\ell_P},
\]
which is the random-walk foam model. At the Holometer ($L=40$ m) this predicts
$\delta L=2.5\times10^{-17}$ m against a displacement sensitivity of order
$10^{-18}$ m, and at LIGO ($L=4$ km) it predicts $2.5\times10^{-16}$ m against
roughly $10^{-19}$ m. The prediction exceeds sensitivity by factors of $25$ and
$2.5\times10^{3}$ respectively. \emph{This version of the framework is excluded
by existing interferometry.}

\paragraph{Area scaling.} If instead $n\sim(L/\ell_P)^2$, as a holographic
counting of Planck-area cells would suggest, then
\begin{equation}
\delta L \;=\; L\left(\frac{\ell_P}{L}\right) \;=\; \ell_P
\label{eq:planckuniversal}
\end{equation}
\emph{exactly, and independently of $L$}. The framework then predicts a
universal, scale-free length uncertainty of precisely one Planck length,
$1.6\times10^{-35}$ m, far below any foreseeable sensitivity but at least a
sharp and unusual statement: the noise does not grow with the size of the
apparatus, unlike in every foam model that current interferometry constrains.

The conclusion is that existing data already select the area law within this
framework. We regard this as the most useful experimental result available
here, since it converts an unfixed modelling choice into one that observation
has decided.

\subsection{What is not observable}

Adopting the bound \eqref{eq:nbound-value}, the remaining predictions are
numerically hopeless, and we state this plainly. With $n=10^{23}$:
\begin{itemize}
\item the rapidity ladder has spacing $\delta\phi\sim1/n\approx10^{-23}$;
\item the fractional interval noise is $n^{-1/2}\approx3\times10^{-12}$, and by
\eqref{eq:planckuniversal} the corresponding length noise is one Planck length;
\item a violation of velocity addition carrying even a modest entropy deficit
$J=10^{-3}$ is suppressed by $e^{-nJ}=e^{-10^{20}}$, which is zero for every
practical and impractical purpose alike.
\end{itemize}
The last item is the price of the resolution obtained in
Section~\ref{sec:maxent}: having established that departures from relativistic
composition are suppressed at a calculable exponential rate, we have thereby
established that they will never be seen.

\subsection{The one loophole: minimum-entropy probes}

There is a single place where the finite-$n$ structure escapes suppression by
$n$, and it follows from Eq.~\eqref{eq:Ng}. The number of measurements needed
to resolve the rapidity ladder is
\[
N\simeq\epsilon(1-\epsilon)\,g,
\]
where $g$ is the number of \emph{blank} positions in the probe map --- those
carrying neither a clock nor a ruler step. This is independent of $n$
altogether. A probe with $g=4$ resolves the ladder with of order one
measurement whatever the sequence length, and the maximal rest map $g=0$ does
so with certainty, its re-reading being exactly deterministic
(Section~\ref{sec:fluct}).

The physical characterisation of such a probe is worth stating, because it is
not merely a formal escape. By the counting of Section~\ref{sec:entropy}, $g=0$
means $\Omega=1$ in the displacement sector: there is exactly one microstate,
and the probe carries zero entropy. The criterion for observing the
discreteness is therefore
\begin{quote}
\emph{use probes of minimal entropy, not probes of high energy.}
\end{quote}
This inverts the usual strategy for quantum-gravity phenomenology, which seeks
the largest available $E/E_{\rm Planck}$. Here energy does not help, because
every transformation is exactly Lorentz invariant at any energy; what helps is
a probe whose displacement sector is maximally constrained.

We do not know whether any physical system realises such a probe, and we make
no claim that one does. What Eq.~\eqref{eq:Ng} provides is a specific property
to look for, and we regard identifying a physical realisation --- or proving
that none exists --- as the most promising direction for making this framework
empirically live.

\subsection{Systems at maximal information capacity}

Throughout the paper $n$ has been a regulator with only a lower bound: cosmic
rays require $n\gtrsim10^{23}$ (Section~\ref{sec:experiment}), and for any
laboratory system the substrate can always be taken larger, so every
finite-$n$ effect can be made invisible by fiat. There is one class of
systems for which this escape is unavailable: those that saturate an
information bound. A black hole horizon and a de Sitter horizon carry, on
holographic grounds, a fixed and finite number of independent binary degrees
of freedom --- the Bekenstein--Hawking entropy in bits --- and if the
sequence length is identified with that capacity, in the spirit of the area
counting $n\sim(L/\ell_P)^2$ that interferometry already selects within this
framework (Eq.~\eqref{eq:planckuniversal}), then $n$ is \emph{pinned}. For
such systems the corrections derived in this paper stop being adjustable
smallness and become parameter-free floors. We therefore agree with the
natural expectation: objects at maximal information capacity are where
deviations from exact relativity should be sought, because they are the only
objects for which the theory cannot hide them.

The magnitudes divide into three regimes. For large horizons the generic
floor is fractional, of order $n^{-1/2}$: about $3\times10^{-39}$ for a
solar-mass black hole ($n\sim10^{77}$) and $7\times10^{-62}$ for the present
de Sitter horizon ($n\sim2\times10^{122}$) --- irreducible in principle,
hopeless in practice. The effects are instead order unity when the capacity
itself is small: a black hole in the final stage of evaporation has
$n\to O(1)$, where the rapidity spectrum \eqref{eq:spectrum} is visibly
discrete, $\gamma_{\max}$ is of order one, and the causal classification is
ambiguous over most of count space --- the framework predicts that special
relativity fails outright at the endpoint, which is where one expects
semiclassical physics to fail anyway. In between sits inflation: a quasi--de
Sitter phase with Hubble rate $H$ has $n=\pi(M_P/H)^2$, so the floor is
$n^{-1/2}=H/(\sqrt\pi\,M_P)$ --- parametrically the amplitude of
horizon-scale fluctuations in de Sitter space. We flag this as a
consistency of scales, not a derivation.

The sharpest realisation of the expectation, however, is local and holds for
horizons of \emph{any} size. The accompaniment bound of
Section~\ref{sec:causal} caps the relative Lorentz factor of any two
admissible frames at $\gamma_{\max}\simeq\sqrt n/2$
(Eq.~\eqref{eq:gammamax}). A static observer hovering at proper distance $d$
from a Schwarzschild horizon has $\gamma\simeq 2r_s/d$ relative to a frame
falling freely from rest at infinity; with the holographic assignment
$n=\pi(r_s/\ell_P)^2$ the bound $\gamma\le\gamma_{\max}$ becomes
\[
d\;\gtrsim\;\frac{4}{\sqrt\pi}\,\ell_P\;\approx\;2\,\ell_P ,
\]
independently of the mass of the hole. Within a Planck proper length of the
horizon there is no admissible frame that can accompany a static worldline:
the static observer exits the accompaniable class, and the framework
produces, from counting, the stretched horizon of black-hole thermodynamics
at its conventional depth. In this local sense the deviations are not merely
largest at maximal-capacity objects; they are $O(1)$ in a Planck-thin layer
at the horizon of every black hole, however large \cite{mp,Dai:2020irc,Lin:2025jmz}.

The identification of $n$ with horizon entropy
is imported from holography, not derived here, and the framework as
constructed is special-relativistic --- flat, with globally defined frames
--- so statements about curved horizons are translations through local
inertial frames, heuristic by construction. Within those limits, the
conclusion stands: capacity-saturating systems are the natural arena for the
finite-$n$ phenomenology of Section~\ref{sec:experiment}, and the smaller
the capacity, the larger the departure.

\subsection{Summary}

\begin{center}
\begin{tabular}{lll}
\toprule
Prediction & Status & Comment \\
\midrule
No energy-dependent dispersion & consistent with data & falsified by any LIV detection \\
No vacuum birefringence & consistent with data & shared with standard SR \\
$\gamma_{\max}=O(\sqrt n)$ & bounds $n\gtrsim10^{23}$ & from UHECR; assumes single map \\
$\delta L=\sqrt{L\ell_P}$ (linear $n$) & \textbf{excluded} & LIGO, Holometer \\
$\delta L=\ell_P$ (area-law $n$) & allowed & universal, scale-free, unobservable \\
Rapidity ladder $\delta\phi\sim1/n$ & unobservable & unless $g$ is small \\
Entropy-suppressed composition & unobservable & $e^{-nJ}$ with $n\gtrsim10^{23}$ \\
Ladder via minimum-entropy probe & open & $N\sim\epsilon(1-\epsilon)g$, $n$-independent \\
\bottomrule
\end{tabular}
\end{center}

The framework is at present constrained rather than
tested. It is falsifiable, one of its variants is already dead, and it makes a
sharp universal prediction \eqref{eq:planckuniversal} that no foreseeable
experiment can reach. Its value is currently conceptual, with possible confirmation/falsification in future.

\section{Discussion}
\label{sec:discussion}

\subsection{What kind of theory this is}

We do not assert that Lorentz transformations are elements of the XOR algebra;
they are not, and Section~\ref{sec:assumptions} says so explicitly. What we
assert is that a counting theory on binary sequences, coarse-grained in the
manner dictated by the counting postulate, has an effective description whose
invariance group is the Lorentz group and whose invariant is the Minkowski
interval.

The analogy with hydrodynamics is worth taking seriously. The Navier--Stokes equations have symmetries that molecular
dynamics does not, and the emergence of those symmetries is not a defect of the
microscopic theory but the ordinary behaviour of coarse-graining. Likewise here:
the substrate is abelian, involutive, and possesses no notion of interval, while
the effective theory is non-abelian, has an invariant cone, and possesses an
indefinite metric. Nothing was smuggled in; the structures appear because
averaging over microstates is a projection, and projections have their own
symmetries. Viewed this way, the cone is not merely compatible with the
counting theory: it is the eigenvector structure of the unique averaging
matrix, and its $\epsilon$-independence is what makes it observer-independent.
This is as close to a derivation of causal structure from non-geometric
premises as we know how to come.

\subsection{The role of irreversibility}

The identification of the conformal factor with entropy production
(Section~\ref{sec:entropy}) seems to us the most suggestive feature of the
framework. In the usual presentation of relativity there is no statistics
anywhere near the kinematics. Here the two are entangled from the start: the
coarse-grained flow is irreversible, the light cone is its entropy maximum, and
Lorentz symmetry is precisely the reversible remainder after the (information) entropy
production has been factored out.

If one takes this seriously, the reason spacetime looks Lorentzian is that
Lorentz transformations are the transformations that lose no information. That
is a statement one could imagine generalising: in a setting where the flow is
not exactly doubly stochastic, one would expect a deformed invariance group,
with the deformation controlled by the entropy production. We have not pursued
this.

\section{Conclusions}
\label{sec:conclusions}

We have formulated the derivation of relativistic kinematics from binary
sequences as a problem in statistical coarse-graining. The chain of reasoning
is: the counting postulate makes symbol positions unphysical; averaging over
positions produces a unique mixing matrix $S(\epsilon)$; that matrix has
$\epsilon$-independent eigendirections, which therefore constitute an
observer-independent causal cone; the unique quadratic form vanishing on the
cone is the Minkowski interval; and the interval-preserving part of the flow
is $SO(1,1)$. At no point is a metric, a light cone, or a Lorentz
transformation assumed.

Four results seem to us worth emphasising. The invariance of the null
coordinate $u=\hat C_M+\hat D_M$ holds \emph{exactly}, for every microscopic configuration
and every finite $n$, so half the light-cone structure is not a coarse-grained
statement at all. The conformal factor that must be removed to obtain a Lorentz
transformation is the entropy production of the coarse-graining, making Lorentz
symmetry the reversible core of an irreversible flow. In $3+1$ dimensions the
index structure of the Poincar\'e algebra follows from reading symbols as
transpositions of registers, so that the non-commutativity of the Lorentz group
is present in the alphabet even though XOR is commutative. And relativistic
velocity addition, which at the level of counts is not an identity, holds for
almost all triples of observers with discrepancy $O(n^{-1/2})$ and exponentially
suppressed exceptions, whose rarity is measured by their entropy
deficit under the flat counting measure the framework already assumes.

That last result determines how the whole construction should be read:
\begin{quote}
\textbf{Lorentz symmetry is the typical large-$n$ behaviour of XOR counting.}
\end{quote}
Two structures usually postulated independently --- the causal cone and the
group law --- turn out to be the same statement twice: the light cone is the
maximum-entropy locus of the count distribution (Section~\ref{sec:entropy}),
and velocity addition is the maximum-entropy composition of frame maps
(Section~\ref{sec:maxent}). Lorentz structure is, in both instances, what
maximises entropy under the counting measure. Velocity addition therefore
holds in this theory in the sense that the second law holds in thermodynamics.
We regard this as the correct logical shape for an emergent-spacetime claim: a
coarse-grained symmetry that held exactly for every microscopic configuration
would not be emergent at all; it would be a symmetry of the substrate wearing
a disguise.

We should also be clear about the division of labour among the inputs. The
counting postulate is doing the greatest share of the work --- it supplies,
in particular, the independence of placements on which the composition law
rests (Section~\ref{sec:maxent}); the product
structure and equal weight together fix the signature; and the interior of
the cone is selected not by a restriction but by the persistence clause of
input (iv) of Section~\ref{sec:inputs}, through the accompaniment mechanism
of Section~\ref{sec:causal}. What XOR contributes on its own
is the observer-independence of maps, the exact cocycle, and the exact
invariance of $u$. A reader who wishes to say that ``the assumptions already
encode a light cone'' is not wrong, and the interesting question is not whether
inputs were used but whether the particular inputs used are more or less
palatable than postulating Lorentz invariance directly. We think a flat measure
over microstates is a mild assumption and a metric is a strong one, but we
recognise that this is a judgement rather than a demonstration.

On the experimental side the situation is as follows. The exactness of the Lorentz
invariance, the framework's chief structural asset, leaves it predicting the
absence of every signal that quantum-gravity phenomenology currently searches
for. Ultra-high-energy cosmic rays already impose $n\gtrsim10^{23}$ through
the single-map ceiling; interferometry has excluded the linear-scaling variant
of the model, selecting the area law and with it a universal length
uncertainty of one Planck length; and the remaining predictions are
numerically out of reach. The one direction that escapes suppression by $n$ is
the minimum-entropy probe of Eq.~\eqref{eq:Ng}, and identifying a physical
realisation of one is in our view the most promising route to making the
framework empirically live. The other direction is to study systems at the maximum of their information capacity, like black holes or de Sitter space.

What remains,  is a construction in which causal
structure, metric signature, and Lorentz invariance are consequences of
counting rather than assumptions about spacetime. Whether that is the beginning
of a physical theory or an instructive mathematical fact about coarse-graining
is not something the present work can settle.

{\bf Acknowledgments}

 Claude and Copilot were used in creation of this manuscript, mainly to test various possible realizations of the framework, generate the concrete examples, numerical checks and improve the text. D.S. is partially supported by the US National Science Foundation, under Grant No.  PHY-2310363.

\appendix

\section{Worked Example: Frame Change at $n=4$}
\label{app:n4}

We work everything out explicitly at $n=4$, the smallest even length for which
the arithmetic is transparent. Sequences are written as strings of the four
symbols, position by position.

\subsection{Setting up}

Let Alice's reference event and the two events of interest be
\[
e^A=AAAA,\qquad e_1=AAAA,\qquad e_2=CCDA .
\]
The worldline map is
\[
M_{12}=e_1\oplus e_2=AAAA\oplus CCDA=CCDA ,
\]
with counts $\hat C_M=2$, $\hat D_M=1$, $\hat A_M=1$, $\hat B_M=0$. Hence
\[
\hat C_M=2,\qquad \hat D_M=1,\qquad u=3,\qquad w=1,
\]
and
\[
ds^2=\hat C_M^2-\hat D_M^2=4-1=3,\qquad v=\frac{\hat D_M}{\hat C_M}=\frac12 .
\]
The map is timelike, $\hat C_M>\hat D_M$, as required for a worldline.

\subsection{The frame map and its rapidity}

Let Bob's reference differ from Alice's by
\[
\beta=BAAA ,
\]
so $\hat B_\beta=1$ and $\epsilon=1/4$. From Eq.~\eqref{eq:rapidity},
\[
\phi=\frac12\ln\frac{n}{n-2\hat B_\beta}=\frac12\ln\frac{4}{2}=\frac12\ln2\approx0.3466,
\]
and correspondingly
\[
v_\beta=\frac{\hat B_\beta}{n-\hat B_\beta}=\frac13,\qquad
\cosh\phi=\frac{3}{2\sqrt2}\approx1.0607,\qquad
\sinh\phi=\frac{1}{2\sqrt2}\approx0.3536 .
\]
At $n=4$ the admissible range is $\hat B_\beta\in\{0,1\}$: at
$\hat B_\beta=2=n/2$ the frame separation would sit exactly on the light cone,
$v=1$. The accessible rapidities are therefore only $\phi=0$ and
$\phi=\tfrac12\ln2$, a spacing of order unity --- an artifact of the tiny
illustrative $n$; by Eq.~\eqref{eq:spacing} the spacing falls as $1/n$.

\subsection{Microscopic re-reading versus the mean}

Bob re-reads $M_{12}=CCDA$ at the positions where $\beta$ carries a $B$, namely
position $1$. Position $1$ of $M$ carries a $C$, so it is re-read as a $D$:
\[
M'=DCDA,\qquad \hat C_{M'}=1,\;\hat D_{M'}=2 .
\]
Check the exact invariance of $u$ noted below Eq.~\eqref{eq:uw}:
$u'=1+2=3=u$, as promised. The
other null coordinate has moved, $w'=1-2=-1$, consistent with
$w'=w-2(k-l)$ with $k=1$, $l=0$.

Had the single $B$ instead sat at position $3$, which carries a $D$, we would
have found $\hat C_{M'}=3$, $\hat D_{M'}=0$, again $u'=3$ but now $w'=3$. Averaging over the four
equally weighted placements of one $B$ among four positions gives
\[
\langle k\rangle=\epsilon \hat C_M=\tfrac12,\qquad
\langle l\rangle=\epsilon \hat D_M=\tfrac14,\qquad
\langle w'\rangle=w-2(\tfrac12-\tfrac14)=\tfrac12=(1-2\epsilon)w ,
\]
which is exactly Eq.~\eqref{eq:S}. Meanwhile Eq.~\eqref{eq:varw} gives
\[
\mathcal F=\frac{1\cdot3}{3}=1,\qquad
\mathrm{Var}(w')=4\Bigl(\frac34-\frac1{16}\Bigr)=\frac{11}{4}=2.75 ,
\]
which is large compared with $\langle w'\rangle^2=1/4$. At $n=4$ the
coarse-grained description is very poor, exactly as the $n^{-1/2}$ scaling
predicts; the mean-field picture is a statement about large $n$.

\subsection{The Lorentz transformation}

Applying Eq.~\eqref{eq:boost} to the mean-transformed vector,
\[
\begin{pmatrix}\hat C_{M'}\\ \hat D_{M'}\end{pmatrix}
=\begin{pmatrix}\cosh\phi&\sinh\phi\\ \sinh\phi&\cosh\phi\end{pmatrix}
\begin{pmatrix}2\\1\end{pmatrix}
=\begin{pmatrix}2(1.0607)+1(0.3536)\\2(0.3536)+1(1.0607)\end{pmatrix}
=\begin{pmatrix}2.4749\\1.7678\end{pmatrix},
\]
and
\[
ds'^2=(2.4749)^2-(1.7678)^2=6.125-3.125=3.000=ds^2 ,
\]
so the interval is preserved exactly, as it must be since
$\cosh^2\phi-\sinh^2\phi=1$ identically. By contrast the unfactorised flow
gives
\[
S(\tfrac14)\begin{pmatrix}2\\1\end{pmatrix}=\begin{pmatrix}1.75\\1.25\end{pmatrix},
\qquad ds'^2=3.0625-1.5625=1.5=(1-2\epsilon)\,ds^2 ,
\]
confirming Eq.~\eqref{eq:conformal} with $1-2\epsilon=1/2$. The difference
between the two is precisely the dilation $\sqrt{1-2\epsilon}=1/\sqrt2$, and
the entropy interpretation of Section~\ref{sec:entropy} applies to it: the
normalised distribution has moved from $(2/3,1/3)$ to $(0.583,0.417)$, with
$H$ rising from $0.6365$ to $0.6792$.

\section{Time Dilation}
\label{app:dilation}

Take a clock at rest in Alice's frame, ticking twice:
\[
M_{\rm clock}=CCAA,\qquad \hat C_M=2,\;\hat D_M=0,\qquad ds^2=4,\qquad v=0 .
\]
The proper time is $\Delta\tau=\sqrt{ds^2}=2$.

In Bob's frame, boosted with $\phi=\tfrac12\ln2$ as above,
\[
\Delta t'=\hat C_M\cosh\phi=2(1.0607)=2.1213,\qquad
\Delta x'=\hat C_M\sinh\phi=2(0.3536)=0.7071 .
\]
Then
\[
ds'^2=(2.1213)^2-(0.7071)^2=4.5-0.5=4.000=ds^2 ,
\]
so the proper time is unchanged while the coordinate time has increased from
$2$ to $2.1213$. The dilation factor is
\[
\frac{\Delta t'}{\Delta\tau}=\frac{2.1213}{2}=1.0607=\cosh\phi=\gamma ,
\]
the standard result: the moving clock runs slow by exactly $\gamma$.

Note that this map is the maximal rest map of Section~\ref{sec:fluct} in
miniature: with $\hat D_M=0$, every $B$ symbol of $\beta$ that lands on the map must
land on a clock position. For the full maximal case $\hat C_M=n$ the re-reading would
be exactly deterministic and $\mathrm{Var}(w')=0$; here $\hat C_M=2<n=4$, so blank
positions remain and some fluctuation survives, in accordance with
Eq.~\eqref{eq:Ng}.

\section{Length Contraction}
\label{app:contraction}

Length contraction requires more care than time dilation, because measuring a
length means locating both ends \emph{simultaneously}, and simultaneity is
frame-dependent. The relevant map is therefore not a worldline but a
spacelike connecting map between two simultaneous endpoints, for which
$\hat C_{\rm map}=0$; it is a rod orbit in the classification of
Section~\ref{sec:causal}, and the time-ordering property derived there
applies to worldlines and not to such slices.

Let a rod at rest in Alice's frame have proper length $L_0=2$, described by
\[
M_{\rm rod}=DDAA,\qquad \hat C_M=0,\;\hat D_M=2 .
\]
In Bob's frame the two ends are no longer simultaneous. Bob must therefore
compare the two endpoints at equal $t'$, which shifts one endpoint along its
worldline by
\[
\Delta t=L_0\tanh\phi=2\left(\frac13\right)=\frac23 ,
\]
and the length he measures is
\[
L=-\Delta t\,\sinh\phi+L_0\cosh\phi
=-\frac23(0.3536)+2(1.0607)=1.8856 .
\]
This equals
\[
\frac{L_0}{\gamma}=\frac{2}{1.0607}=1.8856 ,
\]
so the rod is contracted by exactly $\gamma$, as required. The whole effect
comes from the relativity of simultaneity, which in this framework is the
statement that Bob's re-reading assigns clock and ruler roles to positions
differently than Alice's does.

\section{Inverse Boosts and the Missing Sign}
\label{app:inverse}

Since $\beta_{AB}=\beta_{BA}$, the counts of the frame map are identical
whichever observer is regarded as transforming. The magnitude of the rapidity
is therefore common to both directions,
\[
|\phi_{AB}|=|\phi_{BA}|=\operatorname{artanh}\frac{\hat B_\beta}{n-\hat B_\beta},
\]
while the transformations themselves are inverse to one another,
\[
\Lambda(\phi)\Lambda(-\phi)=\mathbb{I} .
\]
Concretely, with $\phi=\tfrac12\ln2$,
\[
\Lambda(\phi)=\begin{pmatrix}1.0607&0.3536\\0.3536&1.0607\end{pmatrix},
\qquad
\Lambda(-\phi)=\begin{pmatrix}1.0607&-0.3536\\-0.3536&1.0607\end{pmatrix},
\]
and their product is the identity to the digits shown.

The direction of the boost is thus encoded in \emph{which observer applies the
transformation} and not in any property of $\beta$. As emphasised in
Section~\ref{sec:orientation} this is a limitation on what a single frame map
contains, not on the theory: once three or more observers are present the
relational data determine all relative signs
(Section~\ref{sec:orientation}). The present appendix is the two-observer
case, where by construction there is no third observer to supply the relation,
and the residual freedom is therefore maximal.

\section{The Three-Observer Cocycle: A Concrete Failure}
\label{app:cocycle}

We exhibit the count-level failure quantified by
the composition law of Section~\ref{sec:cocycletheorem} with explicit sets,
choosing deliberately
atypical (strongly correlated) configurations of the kind the analysis shows to
be exponentially rare. Take
$n=10$ and let the boost-symbol positions of the two frame maps be
\[
\beta_1:\{1,2,3,4\},\qquad \hat B_1=4,\ \epsilon_1=0.4 ,
\]
and consider three choices for $\beta_2$ with $\hat B_2=2$, $\epsilon_2=0.2$.

\medskip
\noindent\textbf{Disjoint:} $\beta_2:\{5,6\}$. Then
$\beta_{\rm tot}=\{1,2,3,4,5,6\}$, so $\hat B_{\rm tot}=6$ and
$\epsilon_{\rm tot}=0.6>\tfrac12$: the composite is \emph{super-luminal} and no
rapidity exists.

\medskip
\noindent\textbf{Nested:} $\beta_2:\{1,2\}$. Then
$\beta_{\rm tot}=\{3,4\}$, so $\hat B_{\rm tot}=2$ and $\epsilon_{\rm tot}=0.2$,
giving $\phi_{\rm tot}=0.2554$. But $\phi_1+\phi_2=0.8047+0.2554=1.0601$, and
$\phi_1-\phi_2=0.5493$. Neither matches.

\medskip
\noindent\textbf{Independent-placement overlap:} the expected overlap is
$\hat B_1\hat B_2/n=0.8$, which is not an integer at $n=10$; the composition law
\eqref{eq:epstot} gives $\epsilon_{\rm tot}=0.4+0.2-2(0.08)=0.44$ and
$\phi_{\rm tot}=1.0601=\phi_1+\phi_2$, as required.

\medskip
The lesson is that additivity at the level of counts is recovered only when the
overlap takes its mean value. The configurations above were chosen to be
maximally correlated, and they are exactly the configurations
Section~\ref{sec:cocycletheorem} shows to be exponentially suppressed: at $n=10$
such triples are common, while by $n=400$ the probability of a deviation
$\delta=0.05$ has already fallen below $10^{-6}$. Note also that the exact
subset cocycle is untouched throughout --- in every case above
$S_{AB}\triangle S_{BC}=S_{AC}$ holds identically. What fails is only the
projection to counts, which is the content of the composition law.

\section{The Symbol Permutations in $3+1$ Dimensions}
\label{app:perm}

We tabulate the permutation action \eqref{eq:transposition} explicitly. With
registers ordered $(t,x,y,z)$,
\[
B_x=(t\,x),\quad B_y=(t\,y),\quad B_z=(t\,z),\quad
R_x=(y\,z),\quad R_y=(z\,x),\quad R_z=(x\,y).
\]

\paragraph{Non-commutativity.} Take $B_x$ and $R_z$, which share the register
$x$. Acting on $(t,x,y,z)$,
\[
R_z\!\circ\!B_x:\;(t,x,y,z)\mapsto(y,t,x,z),\qquad
B_x\!\circ\!R_z:\;(t,x,y,z)\mapsto(x,y,t,z),
\]
which differ. Of the $36$ ordered pairs, $24$ fail to commute, and the six
symbols generate $S_4$.

\paragraph{Conjugation.} The key computation is
\[
B_x\,B_y\,B_x^{-1}=(t\,x)(t\,y)(t\,x)=(x\,y)=R_z ,
\]
reproducing $[K_x,K_y]\propto J_z$: two boosts in different directions generate
a rotation. Similarly
\[
R_z\,B_x\,R_z^{-1}=(x\,y)(t\,x)(x\,y)=(t\,y)=B_y ,
\]
reproducing $[J_z,K_x]\propto K_y$, and
\[
R_x\,R_y\,R_x^{-1}=(y\,z)(z\,x)(y\,z)=(x\,y)=R_z ,
\]
reproducing $[J_x,J_y]\propto J_z$. All thirty-six conjugates lie inside the
six-symbol set and match the Lie-algebra index pattern with no exceptions.

\paragraph{The missing signs.} Note that conjugation gives
$B_x\triangleright B_y=R_z$ and $B_y\triangleright B_x=R_z$ alike, whereas
$[K_x,K_y]=-[K_y,K_x]$. The permutation picture is blind to the antisymmetry,
which is the algebraic face of the orientation problem of
Appendix~\ref{app:inverse}.

\section{Mirrors, Pairs, and One Bit: Resolving the Rotation Composition Problem}
\label{sec:mirrors}

This appendix carries out in full the programme summarised at the end of
Section~\ref{sec:angles}, pushing the diagnosis given there to a proposed
resolution. The argument has three
steps: a no-go theorem showing that the problem is unsolvable as stated; a
re-reading of what a frame map represents, under which exact composition
follows from an identity the framework already possesses; and a one-bit
extension of the alphabet under which exact composition holds at the symbol
level, with the bit identified as a convention the framework already owes.

\subsubsection*{Step 1: no calibration can work}

Let same-axis rotation counts compose by the independent-placement union rule
$\rho_{\rm tot}=\rho_1+\rho_2-2\rho_1\rho_2$, with $\rho=\hat R_\beta/n$. Then
no calibration $\theta(\rho)$ whatsoever makes this composition realise a
nontrivial compact group; the only elements that compose exactly and
recurrently are $\rho=0$ and $\rho=1$. To see this, write
$u=1-2\rho\in[-1,1]$. The union rule is $u_{\rm tot}=u_1u_2$, so
$|u_{\rm tot}|=|u_1||u_2|$, which is strictly smaller than $|u_1|$ unless
$|u_2|=1$. Hence any element with $0<\rho<1$ generates under repeated
composition a sequence with $|u|^k\to0$ monotonically: it never returns to,
or near, the identity $u=1$. In any group, and a fortiori in any compact
group, the powers of an element of finite order return to the identity and the
powers of an element of infinite order return arbitrarily close to it;
monotone contraction forbids both. The recurrent boundary set $\{u=\pm1\}$ is
the two-element group $\mathbb Z_2$.

This turns the suspicion recorded in Section~\ref{sec:angles} into a proof: the
obstruction is the arithmetic, not the calibration, and the half-angle choice
of Eq.~\eqref{eq:angle} loses nothing, since nothing could win. The surviving
$\mathbb Z_2$ is meaningful: $\rho=1$ is the all-exchange map, calibrated to
$\theta=\pi$, and it composes correctly even with generic elements ---
$\rho_2=1$ gives $\rho_{\rm tot}=1-\rho_1$, i.e.\ $\theta_{\rm tot}=\pi-\theta_1$,
which is $\pi+\theta_1$ up to the undetermined sense of rotation, and that
sense is once again the conventional bit of input~(v) of
Section~\ref{sec:inputs}. Pure counting thus
delivers exactly one nontrivial rotation, the half-turn, and the half-turn is
an involution. This is the clue.

\subsubsection*{Step 2: the arithmetic of counting is split-complex}

Where the obstruction lives can be stated in one sign. Assemble the counts of
a rotation map into the number
\begin{equation}
z_\beta=(n-\hat R_\beta)+\mathrm{j}\,\hat R_\beta ,\qquad \mathrm{j}^2=+1 ,
\label{eq:splitj}
\end{equation}
an element of the \emph{split}-complex plane. A short computation shows that
the union rule is precisely multiplication there,
\[
n\,z_{\rm tot}=z_1z_2 ,
\]
component by component: the real part composes as
$(n-\hat R_1)(n-\hat R_2)+\hat R_1\hat R_2$ and the $\mathrm j$-part as
$\hat R_1(n-\hat R_2)+(n-\hat R_1)\hat R_2$. Complex multiplication would
differ in a single sign, $-\hat R_1\hat R_2$ in the real part. That one sign
is the entire content of the rotation problem. With $\mathrm j^2=+1$ the
unit-modulus elements form hyperbolas, $e^{\mathrm j\phi}=\cosh\phi+\mathrm
j\sinh\phi$, and the split plane contains no circle: this is the geometric
form of the no-go statement above, and it is simultaneously the reason the
\emph{boost} sector works, since the same identity with boost counts reads
$n(n-2\hat B_{\rm tot})=(n-2\hat B_1)(n-2\hat B_2)$, the exact
multiplicativity on which the rapidity calibration rests. XOR counting is
split-complex arithmetic: hyperbolic angles are native to it, circular angles
are not. The missing sign $\mathrm i^2=-1$ also has a name in rotation theory:
it is the statement that a $2\pi$ rotation is not the identity but its
negative --- the spinor sign. Counts are non-negative multiplicities of an
involution, $R\oplus R=A$; they are structurally $\mathrm j^2=+1$.

\subsubsection*{Step 3a: frame maps are mirrors}

An involution cannot represent a generic rotation, but it represents a
reflection exactly, and by the Cartan--Dieudonn\'e theorem every rotation is
the product of two reflections, with rotation angle equal to \emph{twice} the
angle between the mirrors. We therefore propose to read the assignment
$\beta\mapsto$ transformation differently: a rotation frame map is not a
rotation but a \emph{mirror}, whose relational inclination is the
count-calibrated $\theta/2$ of Eq.~\eqref{eq:angle}, and a physical rotation
is an \emph{ordered pair} of frame maps. Under this reading, Eq.~\eqref{eq:angle}
is not the calibration of a rotation at all; it calibrates a mirror, and the
factor $\tfrac12$ is Cartan--Dieudonn\'e's.

Composition is then exact by cancellation. For pairs sharing a middle map,
\begin{equation}
(\beta_3,\beta_2)\circ(\beta_2,\beta_1)=(\beta_3,\beta_1),
\label{eq:pairlaw}
\end{equation}
because the middle mirror acts twice and $\beta\oplus\beta=A^n$
\emph{exactly}. The identity paraded under ``No group isomorphism'' in
Section~\ref{sec:assumptions} as the sharpest failure --- every frame map is
an involution, no boost or rotation is --- is thereby re-read as the engine of
the group law: involutivity is precisely the property a reflection must have.
No independence assumption enters Eq.~\eqref{eq:pairlaw} and no typicality is
invoked. The reading also aligns with Section~\ref{sec:orientation}: a single
frame map is symmetric under exchange of the observers, which is correct for
a mirror, since a mirror is unoriented; the ordered pair carries the sense of
rotation, and reversing the pair inverts it. The relational orientation of
Section~\ref{sec:orientation} is exactly the datum that orders the pair.

In $3+1$ dimensions the mirrors are already tabulated:
Appendix~\ref{app:perm} reads each two-flip symbol as a transposition of register
labels, and a transposition of two ruler axes is the reflection across their
diagonal plane. Products of two such reflections are rotations, and the
composition is the appendix's own permutation calculus: composing the
transpositions $(x\,y)$ and $(y\,z)$ gives the $3$-cycle $(x\,y\,z)$, which
geometrically is the rotation by $120^\circ$ about the cube diagonal
$(1,1,1)$ --- an exact, if finite, piece of $SO(3)$ that symbol composition
already contains. For continuous angles, composing rotations about two
different axes by choosing the shared mirror to be the plane containing both
axes reduces the product of four reflections to a single pair; the induced law
on axes and angles is the classical spherical-triangle construction, which is
quaternion multiplication. Exact $SO(3)$ composition is thus available with
no new structure: the group law is supplied by involutivity, the continuum of
mirror inclinations by the calibrated counts. The union rule and its
$O(\theta^3)$ defect are re-diagnosed as what one gets by composing pairs
that share no middle map and then projecting onto counts under independent
placement: the same exact-for-subsets, typical-for-counts dichotomy that
Section~\ref{sec:cocycletheorem} established for boosts. The rotation sector is
thereby restored to parity with the boost sector, rather than being
materially worse.

\subsubsection*{Step 3b: one bit buys the double cover}

There is a second route, which stays at the level of symbols and quantifies
exactly what the circle costs. The ruler-exchange symbols
$\{A,R_x,R_y,R_z\}$ form a Klein four-group under XOR. Centrally extend it by
a single sign: introduce one new symbol $\bar A$ with $\bar A^2=A$, and lift
each exchange so that it squares to the sign rather than to the identity,
\begin{equation}
R_a^2=\bar A,\qquad
R_xR_y=R_z,\qquad R_yR_x=\bar A\,R_z\quad\text{(cyclic)} .
\label{eq:Q8}
\end{equation}
The extension with these properties is unique: it is the quaternion group
$Q_8=\{A,\bar A,\pm R_x,\pm R_y,\pm R_z\}$, realised concretely by
$R_a\mapsto -\mathrm i\sigma_a$, i.e.\ by the $\pi$-rotations in the doublet
representation of Section~\ref{sec:angles}. One bit --- not a complex Hilbert
space per slot --- converts the abelian Klein group into the quaternion
group, and quaternion multiplication \emph{is} exact $SU(2)$ composition; the
calibrated angles then compose by the $SU(2)$ multiplication displayed in
Section~\ref{sec:angles} with no leading-order caveat, and the split-complex sign
of Eq.~\eqref{eq:splitj} is repaired, since signed agreement counts
$\hat A_+-\hat A_-$ compose with the $-\hat R_1\hat R_2$ that the circle
requires.

The bit is not new physics. What distinguishes $R_xR_y$ from $R_yR_x$ in
Eq.~\eqref{eq:Q8} is a handedness, and Section~\ref{sec:orientation} proved
the framework owes exactly this convention already: input~(v) of
Section~\ref{sec:inputs} spends one
global bit on the sign of $\epsilon_{ijk}$. The extension installs that bit
in the algebra instead of leaving it as a label. In cohomological language,
the obstruction to the rotation group is the class of the central extension
$\mathbb Z_2\to Q_8\to\mathbb Z_2\times\mathbb Z_2$; the theory pays for
$SO(3)$ with a bit it had already spent on orientation. And the purchase
comes with interest: since $R_a^2=\bar A\neq A$ while $R_a^4=A$, the extended
framework distinguishes a $2\pi$ rotation from a $4\pi$ one. It does not
merely recover the rotation group; it recovers its double cover, which is to
say the substrate, thus completed, is naturally a $Spin(3)$ theory and
contains spinors --- consonant with the companion formulation, where the
binary register is an isospin doublet.

\subsubsection*{Status}

We separate what is proved from what is proposed. The no-go statement of
Step~1, the split-complex identity below Eq.~\eqref{eq:splitj}, the
Cartan--Dieudonn\'e half-angle, and the group facts of Eq.~\eqref{eq:Q8} are
established results. The mirror reading of a rotation frame map and the
identification of the extension bit with the orientation convention of
input~(v) of Section~\ref{sec:inputs} are interpretive steps, which we
regard as forced by those results but which a reader may weigh
independently. What remains genuinely open is
the typicality statement that would close the circle: deriving the union rule,
with its $O(n^{-1/2})$ error, as the sign-forgetting coarse-graining of
$Q_8$ (equivalently mirror-pair) composition over unshared pairs, in parallel
with the boost analysis of Section~\ref{sec:cocycletheorem}. We expect the
derivation to be of the same concentration type, but we have not carried it
out.

\section{Detailed Derivations}
\label{app:proofs}

\subsection*{The composition law as a law of large numbers}

This section derives the statement of Section~\ref{sec:cocycletheorem}.
The overlap $I=|S_{AB}\cap S_{BC}|$ is hypergeometric, with
\[
\langle I\rangle=\frac{\hat B_1\hat B_2}{n},\qquad
\mathrm{Var}(I)=\hat B_2\,\frac{\hat B_1}{n}\Bigl(1-\frac{\hat B_1}{n}\Bigr)
\frac{n-\hat B_2}{n-1}=n\,b_1b_2(1-b_1)(1-b_2)\bigl[1+O(n^{-1})\bigr].
\]
By Eq.~\eqref{eq:Btot}, $\epsilon_{\rm tot}=(\hat B_1+\hat B_2-2I)/n$, so
$\langle\epsilon_{\rm tot}\rangle=\bar\epsilon$ exactly and
$\mathrm{Var}(\epsilon_{\rm tot})=4\mathrm{Var}(I)/n^2=O(n^{-1})$. Since
$\phi=-\tfrac12\ln(1-2\epsilon)$ has $d\phi/d\epsilon=(1-2\epsilon)^{-1}$,
propagation of the fluctuation gives Eq.~\eqref{eq:sigmadelta}. The mean
discrepancy follows from the second derivative,
$d^2\phi/d\epsilon^2=2(1-2\epsilon)^{-2}$, giving
$\langle\Delta_n\rangle=\mathrm{Var}(\epsilon_{\rm tot})(1-2\bar\epsilon)^{-2}
+O(n^{-2})=O(n^{-1})$. For Eq.~\eqref{eq:largedev}, Hoeffding's inequality for
sampling without replacement gives
$\Pr(|I-\langle I\rangle|\ge tn)\le2\exp(-2t^2n^2/\hat B_2)\le2\exp(-2t^2n)$,
and a deviation $\delta$ in $\phi$ corresponds to a deviation
$t=O(\delta)$ in $I/n$ for $\bar\epsilon$ bounded away from $\tfrac12$.

\subsection*{Signed rapidities from unsigned magnitudes}

This section derives the reconstruction statement of
Section~\ref{sec:orientation}. The stated conditions say exactly that the $d_{ij}$ are the pairwise distances
of $N$ points on a line. Fix any observer $O$ and set $\psi_O=0$; this uses the
additive freedom and is the statement that no frame is preferred. Choose one
further observer $R$ and declare $\psi_R>0$; this uses the reflection freedom.
For any other $X$, the two possibilities $\psi_X=\pm d_{OX}$ are distinguished
by the already-known distance $d_{RX}$, since
\[
d_{RX}=\bigl|d_{OR}-d_{OX}\bigr| \iff X \text{ lies on the same side of } O
\text{ as } R,
\qquad
d_{RX}=d_{OR}+d_{OX} \iff \text{ the opposite side},
\]
and these two values coincide only in degenerate cases $d_{OR}=0$ or
$d_{OX}=0$. Every sign is therefore fixed. Reversing the choice at $R$ maps
$\psi\to-\psi$ throughout.

\section{Summary Table of Counting Dictionary}
\label{app:dictionary}

\begin{center}
\begin{tabular}{ll}
\toprule
Counting quantity & Emergent meaning \\
\midrule
$n$ & sequence length; sets all finite-size effects \\
$\hat C_M,\hat D_M$ & time and space displacement $\Delta t,\Delta x$ \\
$u=\hat C_M+\hat D_M$ & null coordinate; \emph{exactly} invariant \\
$w=\hat C_M-\hat D_M$ & conjugate null coordinate; contracts under the flow \\
$w=0$ & the light cone; entropy maximum of the flow \\
$uw$ & Minkowski interval $ds^2$ \\
$\hat B_\beta$ & boost parameter of a frame change \\
$\epsilon=\hat B_\beta/n$ & mixing fraction \\
$1-2\epsilon$ & conformal factor; entropy production \\
$\tfrac12\ln\frac{1}{1-2\epsilon}$ & rapidity $\phi$ \\
$\hat R_{\beta,i}$ & rotation parameter about axis $i$ \\
$\Omega=n!/\prod\hat X!$ & number of microstates behind a macrostate \\
$\mathrm{Var}(w')$ & interval fluctuation; the finite-$n$ signature \\
\bottomrule
\end{tabular}
\end{center}

\end{document}